\documentstyle[12pt,bezier,axodraw]{article}
 
\makeatletter
 
\def\section{\@startsection {section}{1}{\z@}{+3.0ex plus +1ex minus
  +.2ex}{2.3ex plus .2ex}{\normalsize\bf}}
\def\subsection{\@startsection{subsection}{2}{\z@}{+2.5ex plus +1ex
minus +.2ex}{1.5ex plus .2ex}{\normalsize\bf}}
\def\subsubsection{\@startsection{subsubsection}{3}{\z@}{+3.25ex plus
 +1ex minus +.2ex}{1.5ex plus .2ex}{\normalsize\bf}}

\def\thesection{\Roman{section}}

\def\appendix{\setcounter{section}{0} \setcounter{subsection}{0}
              \setcounter{equation}{0}
	      \@addtoreset{equation}{section}
              \def\thesection{\Alph{section}}
              \def\theequation{\thesection\arabic{equation}}}

\newcount\@tempcntc
\def\@citex[#1]#2{\if@filesw\immediate\write\@auxout{\string\citation{#2}}\fi
  \@tempcnta\z@\@tempcntb\m@ne\def\@citea{}\@cite{\@for\@citeb:=#2\do
    {\@ifundefined
       {b@\@citeb}{\@citeo\@tempcntb\m@ne\@citea
        \def\@citea{,\penalty\@m\ }{\bf ?}\@warning
       {Citation `\@citeb' on page \thepage \space undefined}}%
    {\setbox\z@\hbox{\global\@tempcntc0\csname
b@\@citeb\endcsname\relax}%
     \ifnum\@tempcntc=\z@ \@citeo\@tempcntb\m@ne
       \@citea\def\@citea{,\penalty\@m}
       \hbox{\csname b@\@citeb\endcsname}%
     \else
      \advance\@tempcntb\@ne
      \ifnum\@tempcntb=\@tempcntc
      \else\advance\@tempcntb\m@ne\@citeo
      \@tempcnta\@tempcntc\@tempcntb\@tempcntc\fi\fi}}\@citeo}{#1}}

\def\@citeo{\ifnum\@tempcnta>\@tempcntb\else\@citea
  \def\@citea{,\penalty\@m}%
  \ifnum\@tempcnta=\@tempcntb\the\@tempcnta\else
   {\advance\@tempcnta\@ne\ifnum\@tempcnta=\@tempcntb \else
\def\@citea{--}\fi
    \advance\@tempcnta\m@ne\the\@tempcnta\@citea\the\@tempcntb}\fi\fi}

\makeatother

\expandafter\ifx\csname mathrm\endcsname\relax\def\mathrm#1{{\rm #1}}\fi

\def\nn{\nonumber}
\def\nl{\nonumber\\}

\newcommand{\quabla}{\raisebox{-.2em}{\Large$\Box$}}

\def\beq{\begin{equation}}
\def\eeq{\end{equation}}
\def\beqar{\begin{eqnarray}}
\def\eeqar{\end{eqnarray}}
\def\barr#1{\begin{array}{#1}}
\def\earr{\end{array}}

\def\text{\textstyle}

\def\bma{\begin{displaymath}}
\def\ema{\end{displaymath}}

\def\Ga{\Gamma}
\def\ga{\gamma}
\def\de{\delta}
\def\De{\Delta}

\def\veps{\varepsilon}

\def\si{\sigma}
\def\Si{\Sigma}
 
\def\theenumi{\roman{enumi}}
\def\p@enumi{\theenumi}

\def\refeq#1{\mbox{(\ref{#1})}}
\def\refeqs#1{\mbox{(\ref{#1})}}
\def\refeqf#1{\mbox{(\ref{#1})}}
\def\refse#1{\mbox{Sect.~\ref{#1}}}

\def\refapp#1{\mbox{App.~\ref{#1}}}
\def\citere#1{\mbox{Ref.~\cite{#1}}}
\def\citeres#1{\mbox{Refs.~\cite{#1}}}
 
\newcommand{\ri}{{\mathrm{i}}}

\renewcommand{\d}{{\mathrm{d}}}
\newcommand{\rd}{{\mathrm{d}}}

\renewcommand{\O}{{\cal{O}}}
\renewcommand{\L}{{\cal{L}}}
 
\def\mathswitchr#1{\relax\ifmmode{\mathrm{#1}}\else$\mathrm{#1}$\fi}
\newcommand{\Pf}{\mathswitch  f}
\newcommand{\PW}{\mathswitchr W}
\newcommand{\PZ}{\mathswitchr Z}

\newcommand{\PA}{\mathswitchr A}
\newcommand{\PH}{\mathswitchr H}
\newcommand{\PV}{\mathswitchr V}

\newcommand{\FW}{\mathswitch W}
\newcommand{\FZ}{\mathswitch Z}
\newcommand{\FA}{\mathswitch A}

\def\mathswitch#1{\relax\ifmmode#1\else$#1$\fi}

\newcommand{\MW}{\mathswitch {M_\PW}}

\newcommand{\MZ}{\mathswitch {M_\PZ}}
\newcommand{\MH}{\mathswitch {M_\PH}}
\newcommand{\MV}{\mathswitch {M_\PV}}
 
\newcommand{\rw}{{\mathrm{W}}}
\newcommand{\sw}{\mathswitch {s_{\scriptscriptstyle\mathswitchr W}}}
\newcommand{\cw}{\mathswitch {c_{\scriptscriptstyle\mathswitchr W}}}

\newcommand{\vf}{\mathswitch {v_\Pf}}
\newcommand{\af}{\mathswitch {a_\Pf}}

\def\ie{i.e.\ }

\newcommand{\ses}{self-energies}

\def\OVT{\mbox{o.v.t.}}

\def\RW{R^{1/2}_{\What}}
\def\RWW{R_{\What}}

\def\fbar{\bar f}

\def\Fhat{\hat F}
\def\Vhat{{\hat V}}
\def\Vhatdag{{\hat V}{}^\dagger}
\def\What{{\hat W}}

\def\Zhat{{\hat Z}}
\def\Ahat{{\hat A}}

\def\Hhat{\hat H}

\def\phihat{\hat \phi}
\def\phihatdag{{\hat\phi}^\dagger}
\def\chihat{\hat \chi}
\def\xihat{\hat \xi}
\def\thetahat{\hat\theta}
\def\SF{J_{\Fhat}}
\def\SFc{J_{\Fhat^\dagger}}

\def\SWpm{J_{\What^\pm}}
\def\SWmp{J_{\What^\mp}}
\def\SZ{J_{\Zhat}}
\def\SA{J_{\Ahat}}
\def\SH{J_{\Hhat}}

\def\Sphipm{J_{\phihat^\pm}}
\def\Sphimp{J_{\phihat^\mp}}
\def\Schi{J_{\chihat}}
\def\Sf{J_f}
\def\Sfbar{J_{\fbar}}
\def\Sfpm{J_{f_\pm}}
\def\Sfmp{J_{f_\mp}}
\def\Sfbarpm{J_{\fbar_\pm}}
\def\Sfbarmp{J_{\fbar_\mp}}

\def\xiWhat{\hat\xi_W}
\def\xiAhat{\hat\xi_A}
\def\xiZhat{\hat\xi_Z}

\def\rT{{\mathrm{T}}}

\def\rL{{\mathrm{L}}}

\newcommand{\BF}{\mathrm{BF}}
\newcommand{\GF}{\mathrm{GF}}
\newcommand{\full}{{\mathrm{full}}}

\newcommand{\conn}{{\mathrm{connected}}}

\renewcommand{\Re}{\mathop{\mathrm{Re}}}
 
\newcommand{\Zc}{Z_{\mathrm{c}}}
\newcommand{\Op}{O}
\newcommand{\dgd}[1]{\frac{\de\Ga^\full}{\de#1}}
\newcommand{\dzd}[1]{\frac{\de Z_{\mathrm{c}}}{\ri\de#1}}
\newcommand{\dd}[1]{\frac{\de }{\ri\de#1}}
\makeatletter

\newtoks\@stequation

\def\subequations{\refstepcounter{equation}%
  \edef\@savedequation{\the\c@equation}%
  \@stequation=\expandafter{\theequation}
  \edef\@savedtheequation{\the\@stequation}
  \edef\oldtheequation{\theequation}%
  \setcounter{equation}{0}%
  \def\theequation{\oldtheequation\alph{equation}}}%

\def\endsubequations{%
  \setcounter{equation}{\@savedequation}%
  \@stequation=\expandafter{\@savedtheequation}%
  \edef\theequation{\the\@stequation}\global\@ignoretrue}

\makeatother

\newcommand{\mpar}[1]{{\marginpar{\hbadness10000%
                      \sloppy\hfuzz10pt\boldmath\bf#1}}%
                      \typeout{marginpar: #1}\ignorespaces}

\def\draftdate{\relax}
\def\mda{\relax}
\def\mua{\relax}
\def\mla{\relax}
\def\draft{
\def\thtystars{******************************}
\def\sixtystars{\thtystars\thtystars}
\typeout{}
\typeout{\sixtystars**}
\typeout{* Draft mode!
         For final version remove \protect\draft\space in source file *}
\typeout{\sixtystars**}
\typeout{}
\def\draftdate{\today}
\def\mlas{\vspace*{5mm}\marginpar[\boldmath\hfil$\rightarrow$]%
                   {\boldmath$\leftarrow $\hfil}\vspace*{-5mm}%
                    \typeout{marginpar: $\leftrightarrow$}\ignorespaces}
\def\mdas{\vspace*{3mm}\mpar{$\downarrow$\hfil}\vspace*{-3mm}\ignorespaces}
\def\muas{\vspace*{-5mm}\mpar{$\uparrow$\hfil}\vspace*{5mm}\ignorespaces}
\def\mla{\marginpar[\boldmath\hfil$\rightarrow$\hfil]%
                   {\boldmath\hfil$\leftarrow $\hfil}%
                    \typeout{marginpar: $\leftrightarrow$}\ignorespaces}
\def\mua{\marginpar[\boldmath\hfil$\uparrow$]%
                   {\boldmath$\uparrow$\hfil}%
                    \typeout{marginpar: $\uparrow$}\ignorespaces}
\def\mda{\marginpar[\boldmath\hfil$\downarrow$]%
                   {\boldmath$\downarrow$\hfil}%
                    \typeout{marginpar: $\downarrow$}\ignorespaces}
\def\mla{\marginpar[\boldmath\hfil$\rightarrow$]%
                   {\boldmath$\leftarrow $\hfil}%
                    \typeout{marginpar: $\leftrightarrow$}\ignorespaces}
\overfullrule 5pt
\oddsidemargin -15mm
\marginparwidth 29mm
\marginparpush 2pt
}

\begin{document}
 
\thispagestyle{empty}
\def\thefootnote{\fnsymbol{footnote}}
\setcounter{footnote}{1}
\null
\renewcommand{\baselinestretch}{1}
\Huge
\normalsize
\mbox{} \hfill BI-TP. 96/02\\
\mbox{} \hfill WUE-ITP-96-002\\
\mbox{} \hfill hep-ph/9603341 
\vskip 1cm
\vfill
\begin{center}
{\Large \bf 
Dyson summation without violating Ward identities and 
the Goldstone-boson equivalence theorem
\par} \vskip 2.5em
{\large {\sc Ansgar Denner} \\[1ex]
{\normalsize \it Institut f\"ur Theoretische Physik, Universit\"at W\"urzburg\\
Am Hubland, D-97074 W\"urzburg, Germany}
\\[2ex]
{\sc Stefan Dittmaier} \\[1ex]
{\normalsize \it Theoretische Physik, Universit\"at Bielefeld\\ 
Universit\"atsstra{\ss}e, D-33501 Bielefeld, Germany}
\par} \vskip 1em
\end{center} \par
\vskip 2cm 
\vfil
{\bf Abstract:} \par
In contrast to the conventional treatment of gauge theories,
in the background-field
method the Ward identities for connected Green functions are not violated by
Dyson summation of \ses\ in finite orders of perturbation theory.
Thus, Dyson summation does not spoil gauge cancelations at high energies
which are ruled by the Goldstone-boson equivalence theorem. 
Moreover, in the back\-ground-field method
the precise formulation of the equivalence theorem in higher orders 
(including questions of renormalization)
is simplified rendering actual calculations easier.
Finally, the equivalence theorem is also formulated for the Standard
Model with a non-linearly realized scalar sector and for the gauged
non-linear $\si$-model.
\par
\vfill
\noindent BI-TP. 96/02 \par
\noindent  WUE-ITP-96-002\par
\vskip .15mm
\noindent March 1996 \par
\null
\setcounter{page}{0}
\clearpage
\def\thefootnote{\arabic{footnote}}
\setcounter{footnote}{0}
 
\section{INTRODUCTION}

In many field-theoretic applications, such as the treatment of finite-%
width effects or running couplings, it is desirable or even mandatory
to resum reducible self-energy effects. The use of Dyson resummation
necessarily amounts to an 
incomplete inclusion of higher-order effects, i.e.\ of those which go 
beyond the inspected order of perturbation theory, in theoretical predictions. 
However, it is well known that in the conventional approach to gauge-field 
theories these higher-order effects 
in general violate the Ward identities 
which follow from gauge invariance. These 
rule, in particular, the gauge cancelations 
for longitudinally polarized gauge bosons at high energies.
In order to keep theoretical uncertainties under control, it is 
necessary to preserve the Ward identities exactly
in any finite order of perturbation theory.
In this paper we show that the Ward identities are not violated by
Dyson summation if the gauge theory is quantized
in the framework of the background-field method (BFM).

The discussion of Ward identities is naturally connected to the
investigation of the Goldstone-boson equivalence theorem (ET) 
\cite{Co74,Ch85,Go86,Ya88,Ba90,He92}
which controls the gauge cancelations for $S$-matrix elements.
The ET relates amplitudes for reactions involving
longitudinal vector bosons at high energies to those involving the associated 
would-be Goldstone bosons. 
The gauge cancelations that occur for external
longitudinal gauge bosons are absent for the corresponding scalars.
As the amplitudes for external scalars are much easier to 
evaluate, the ET facilitates the calculation of cross-sections for
reactions with longitudinal vector bosons at high energies
in the Standard Model (SM) and other models. 
But the ET is not only a calculational tool.
Because it relates longitudinal vector bosons to the Higgs
sector, it might allow to derive
information on the mechanism of spontaneous symmetry breaking from the
experimental study of longitudinal vector bosons.
In \citere{He95} it was even pointed out that the validity of the ET itself 
can serve as criterion to discriminate processes for probing 
the electroweak symmetry-breaking sector.

We start by recapitulating
the formulation of the ET as described in the literature.
The ET was derived in the SM for tree-level
amplitudes a long time ago by Cornwall, Levin and Tiktopoulos
\cite{Co74} and extended to all orders in \citeres{Ch85,Go86}.
The derivation of the ET consists of three basic steps:
\begin{enumerate}
\item 
The BRS invariance of the underlying theory implies 
\beq \label{WIET}
\langle A|{\cal T}F_{a_1}(x_1)\cdots F_{a_k}(x_k)|B\rangle_\conn=0 
\eeq
for arbitrarily many insertions of $R_\xi$ gauge-fixing terms 
$F_{a_i} = \partial V_{a_i} + \xi_a c_a \phi_{a_i}$
taken between physical states $|A\rangle$ and $|B\rangle$. 
These identities relate the ``scalar components'' $\partial V_{a_i}$ of 
massive vector fields $V_{a_i}$ with their would-be Goldstone-boson
partners $\phi_{a_i}$ for 
the corresponding external field points $x_i$. 
\label{etstep1}
\item 
The longitudinal polarization vector for high-energetic vector bosons 
with momentum $k_i$ is given by 
$\veps^\mu_{\rL,a_i}(k_i)=k^\mu_i/M_{a_i}+{\cal O}(M_{a_i}/k_i^0)$.
Thus, at high energies 
$\veps_{\rL} V\approx k V \leftrightarrow\partial V$ 
and the identities \refeq{WIET} yield linear relations
between Green functions for longitudinally polarized gauge bosons and their 
would-be Goldstone-boson partners.
Amputation of the ``gauge-fixing legs'' yields relations between the
corresponding transition matrix elements.
For the precise formulation of these relations 
the Ward-Takahashi identities for gauge-boson propagators 
have to be investigated. 
It turns out that in higher orders the simple tree-level form of the 
ET in general is modified by correction factors \cite{Ya88,Ba90,He92},
which depend on the particular choice of the renormalization scheme
in the (unphysical) scalar sector.
Upon exploiting the Ward-Takahashi identities for the propagators, the
correction factors can be expressed in terms of gauge-boson and scalar
self-energies \cite{He92}. By adjusting the
renormalization of the scalar sector appropriately, 
the correction factors can be
absorbed into the renormalization constants \cite{Ba90,He92}.
\item 
Finally, it has to be clarified to which order in $k^0_i/M$
the relation between amplitudes involving 
longitudinal vector bosons and those involving 
would-be Goldstone bosons is valid. 
As far as the SM in the high-energy limit is
concerned \cite{Co74,Ch85,Go86}, unitarity ensures that the scalar
amplitudes coincide with the corresponding longitudinal vector-boson
amplitudes up to ${\cal O}(M/k^0_i)$,
where $M$ generically represents all particle masses present in the SM.
This requires, in particular, that the energy $E$ of the process has to
exceed the Higgs-boson mass $\MH$, i.e.\ $E\gg\MH$.
For arbitrary $\MH$ the leading powers in 
Higgs-boson mass $\MH$ and energy $E$ 
(i.e.\ terms of the order $E^m\MH^n$), 
and thus the range of validity of the ET, can be
determined by power counting for each Feynman graph as described in
\citeres{Ve90,Gr95b}.
\end{enumerate}
Apart from the SM, the ET was also established for chiral Lagrangians
\cite{He94} and the gauged non-linear $\si$-model \cite{Gr95b} in higher
orders. The validity of the ET in the case of general effective
vector-boson interactions at tree level was investigated by power
counting in \citere{Gr95a}.

In this paper we discuss 
the Ward identities for connected Green functions and
the ET for higher orders within the BFM.
The BFM \cite{BFMref,Ab81} allows the construction of a 
gauge-invariant effective action that leads to the same 
$S$ matrix as the conventional effective
action \cite{Ab83}. The resulting vertex and Green functions obey simple
tree-level-like Ward identities even after renormalization \cite{bgflong,Li94}.
We derive these Ward identities for the generating
functional of renormalized connected Green functions. 
A careful amputation procedure leads to identities for amputated Green
functions which imply the ET.
The absence of ghost contributions in the BFM Ward identities simplifies
the precise formulation of the ET as compared to the conventional formalism.
The correction factors to the na\"\i ve ET can be easily obtained
from the transverse parts of the gauge-boson self-energies.

The paper is organized as follows:
In \refse{se:wi} we derive the Ward identities for the generating functional
of connected Green functions within the BFM and give
the Ward identities for on-shell amputated, connected Green functions.
In \refse{se:physsta} we derive the  wave-function renormalization
constants needed for the calculation of $S$-matrix elements. 
The ET is derived in \refse{se:et}
and generalized to the non-linearly realized scalar sector of the SM in
\refse{se:nlscalar}. Section \ref{se:sum} contains a summary and
conclusions. 
In \refapp{app:momconv} we list some conventions.
In \refapp{app:WIproof} we present
the proof of the Ward identities relevant for the ET.
Appendix \ref{app:nomfac} provides the explicit one-loop
results for the wave-function renormalization constants.

\section{WARD IDENTITIES}
\label{se:wi}

\subsection{Background-field effective action}

As we will base our investigations on the BFM it is useful to sketch the
essential ingredients of this approach \cite{BFMref}. We follow 
the treatment of Abbott \cite{Ab81}. The BFM is a technique for
quantizing gauge theories that leads to a gauge-invariant effective action.
To this end the usual fields $\hat\phi$ appearing in the classical 
Lagrangian $\L_{\mathrm{classical}}$ are decomposed into background fields 
$\hat\phi$ 
(marked by a hat) 
and quantum fields $\phi$,
\beq
\label{eq:quantLc}
{\cal L}_{\mathrm{classical}}(\hat\phi) \rightarrow
{\cal L}_{\mathrm{classical}}(\hat\phi + \phi).
\eeq
While the background fields are treated as external sources, only
the quantum fields are variables of integration in the path integral.
The gauge-fixing term with associated (quantum) gauge parameters 
$\xi_Q$,
which is added to allow for the construction of quantum-field propagators, 
is chosen such that the path integral is invariant with respect to
gauge transformations of the background fields. By the usual Legendre 
transformation of the generating functional of
connected Green functions 
with respect to the quantum fields
one obtains an effective action. Putting the
natural sources of this effective action,
which are related to the quantum fields, to zero,
one arrives at an effective action $\Ga[\hat\phi]$
which only depends on background fields. 
This effective action is gauge-invariant, i.e.\ invariant under
gauge transformations of the background fields, which act as sources, and
leads to the same $S$ matrix as the conventional effective action \cite{Ab83}.
The BFM was worked out for the electroweak SM in \citere{bgflong}.
In this reference all Feynman rules including the relevant one-loop
counterterms are listed.
In the following we need only the fact that a gauge-invariant 
background-field effective action $\Ga[\hat\phi]$ exists. 

The invariance of the background-field effective action $\Ga$ under
background-field gauge transformations with associated group parameters
$\thetahat^a$, $a=\FA, \FZ, \pm$,
gives rise to simple tree-level-like
Ward identities:
\beq\label{eq:WIGamma}
\frac{\de\Ga}{\de\thetahat^a} = 0.
\eeq
The explicit form of these functional identities was 
given in \citere{bgfring}
and can also be easily inferred from the results of \citere{bgflong}.
When appropriately renormalized 
\cite{bgflong}, these Ward identities hold 
for the renormalized effective action as well. To this end the field
renormalization constants of the gauge-boson and scalar fields 
must be related to the parameter renormalization constants.
The latter can still be chosen arbitrarily; in particular,
the usual 
on-shell scheme can be adopted for the parameters.
However, since the renormalized fields mix on-shell and 
have propagators with
residues different from one, a 
non-trivial wave-function
renormalization is required when calculating $S$-matrix elements.
In the following, all quantities 
have to be understood as renormalized in the way 
described in \citere{bgflong}%
\footnote{Alternatively, one can avoid a non-trivial wave function
renormalization by introducing appropriate renormalized fields
if one allows for modifications of the renormalized Ward
identities \cite{ek}.\label{foot1}}.

\subsection{Connected Green functions}

The connected Green functions and the $S$ matrix are constructed 
by forming trees with vertices 
from $\Gamma$ connected by lowest-order background-field propagators
\cite{Ab83}. 
In order to define these propagators, one has to add a gauge-fixing term for
the background fields to $\Ga$ resulting in
\beq \label{eq:Gammafull}
\Ga^\full = \Ga + \ri\int\d^4x\, \L^{\BF}_{\GF}.
\eeq
We choose the usual 't Hooft gauge-fixing term
\beq
\L^{\BF}_{\GF} = 
-\frac{1}{2\xiAhat}\left(\partial\hat\FA\right)^2 
-\frac{1}{2\xiZhat}\left(\partial\hat\FZ-\xiZhat\MZ\hat\chi\right)^2 
-\frac{1}{\xiWhat}\left|\partial\hat\FW^+
	-\ri\xiWhat\MW\hat\phi^+\right|^2
\label{eq:bfgf}
\eeq
with the background-gauge-fixing parameters $\xiAhat,\xiZhat,\xiWhat$.
Our conventions for fields, vertex functions, etc.\ follow the ones of
\citere{bgflong} throughout.

{}From the Ward identities \refeq{eq:WIGamma} 
for $\Ga$
one obtains the Ward identities for $\Ga^{\full}$:
\beq\label{eq:WIGammafull}
\frac{\de\Ga^\full}{\de\thetahat^a} = \frac{\de}{\de\thetahat^a} 
\ri\int\d^4x\, \L^{\BF}_{\GF}.
\eeq
The left-hand side is formally identical to the left-hand side of
\refeq{eq:WIGamma} with $\Ga$ replaced by $\Ga^\full$, 
and thus can be simply read off from \citere{bgfring}.
The right-hand side can be evaluated from the behavior of the fields 
in \refeq{eq:bfgf}
under background gauge transformations \cite{bgflong} resulting in 
\beqar \label{eq:dLGF}
\lefteqn{ \frac{\de}{\de\thetahat^\FA} \;\ri \int\d^4x \L^{\BF}_{\GF} =
-\frac{\ri}{\xiAhat}\quabla\partial\Ahat 
+\sum_\pm\frac{(\pm e)}{\xiWhat}
   \left( \What^\pm_\mu\partial^\mu \pm \ri\xiWhat\MW\phihat^\pm\right)
   \left( \partial\What^\mp         \pm \ri\xiWhat\MW\phihat^\mp\right), }
\qquad \nn\\[.5em]
\lefteqn{ \frac{\de}{\de\thetahat^\FZ} \;\ri \int\d^4x \L^{\BF}_{\GF} =
-\frac{\ri}{\xiZhat}\left(\quabla+\xiZhat\MZ^2\right)
\left( \partial\Zhat - \xiZhat\MZ\chihat \right) 
- \frac{\ri e\MZ}{2\cw\sw}\Hhat
   \left( \partial\Zhat - \xiZhat\MZ\chihat \right) }
\qquad\nl &&{}
-\sum_\pm\frac{(\pm e)}{\xiWhat}
   \left( \frac{\cw}{\sw}\What^\pm_\mu\partial^\mu 
   \pm \ri\xiWhat\MW\frac{\cw^2-\sw^2}{2\cw\sw}\phihat^\pm \right)
   \left( \partial\What^\mp \pm \ri\xiWhat\MW\phihat^\mp \right), 
\nn\\[.5em] 
\lefteqn{ \frac{\de}{\de\thetahat^\pm} \;\ri \int\d^4x \L^{\BF}_{\GF} =
  -\frac{\ri}{\xiWhat}\left(\quabla+\xiWhat\MW^2\right)
   \left( \partial\What^\mp \pm \ri\xiWhat\MW\phihat^\mp \right) }
\qquad\nl &&{}
\mp\frac{e}{\xiWhat}\biggl[ 
   \biggl( \Ahat_\mu-\frac{\cw}{\sw}\Zhat_\mu \biggr)\partial^\mu
   \pm \ri\xiWhat\frac{\MW}{2\sw}\left(\Hhat\pm\ri\chihat\right) \biggr]
   \left( \partial\What^\mp \pm \ri\xiWhat\MW\phihat^\mp \right)
\nl &&{}
\pm \frac{e}{\xiAhat}\What^\mp_\mu \partial^\mu\partial\Ahat
\mp \frac{e}{\xiZhat}
   \biggl( \frac{\cw}{\sw}\What^\mp_\mu\partial^\mu
   \mp \ri\xiZhat\frac{\MZ}{2\sw}\phihat^\mp \biggr)
   \left( \partial\Zhat - \xiZhat\MZ\chihat \right),
\hspace{5em}
\eeqar
where $\cw = \MW/\MZ$, $\sw^2=1-\cw^2$, and $e$ denotes the elementary
charge.

The generating functional of connected Green functions, $\Zc$, is obtained 
from $\Ga^\full$ as usual by
a Legendre transformation,
\beq\label{eq:Zc}
\Zc[\SF,\Sf,\Sfbar] = \Ga^\full[\Fhat,f,\fbar] 
+ \ri \int\d^4x \Bigl[\sum_{\Fhat} \SFc\Fhat 
+ \sum_f (\fbar \Sf  + \Sfbar f) \Bigr]
\hspace{2em}
\eeq
with
$\hat F=\Ahat,\Zhat,\What^+,\What^-,\Hhat,\chihat,\phihat^+,\phihat^-$
and
\beqar\label{eq:GtoJ}
\ri \SFc = -\dgd{\hat F}, \qquad 
\ri J_{\fbar} = \dgd{f}, \qquad 
\ri J_f = -\dgd{\fbar}, 
\eeqar
and conversely
\beq\label{eq:ZtoF}
\dzd{\SFc} = \hat F, \qquad
\dzd{J_{\fbar}} = f, \qquad 
\dzd{J_f} = -\fbar.
\eeq
The field $\hat F^\dagger$ 
denotes the complex conjugate of $\hat F$, i.e.\
for instance $\Ahat^\dagger=\Ahat$ but $(\What^+)^\dagger=\What^-$.

As a consequence,
the 1-particle reducible Green functions and $S$-matrix elements are composed
as in the conventional formalism from a tree structure of vertex
functions. While the vertices in these trees are directly given by
the background-field vertex functions, the propagators are determined 
as the inverse of the two-point-vertex functions resulting from $\Ga^{\full}$.

Inserting \refeqf{eq:GtoJ} and \refeqs{eq:ZtoF} into 
\refeqf{eq:WIGammafull} and \refeqf{eq:dLGF}, we find the Ward identities for
the generating functional of connected Green functions in the BFM,
\beqar\label{eq:WIZ}
\lefteqn{ \ri\partial\SA
   + \sum_\pm (\pm e) \dzd{\SWpm^\mu}{\SWpm^\mu} 
   + \sum_\pm (\pm e) \dzd{\Sphipm}{\Sphipm}     
   - e\sum_f Q_f\biggl(\dzd{\Sf}\Sf + \Sfbar \dzd{\Sfbar} \biggr) }
\quad\nl &=&
- \frac{\ri}{\xiAhat}\quabla\partial^\mu\dzd{\SA^\mu} 
+ \sum_\pm \frac{(\pm e)}{\xiWhat}
   \biggl( \dzd{\SWmp^\mu}\partial^\mu \pm \ri\xiWhat\MW\dzd{\Sphimp} \biggr)
   \biggl( \partial^\nu\dzd{\SWpm^\nu} \pm \ri\xiWhat\MW\dzd{\Sphipm} \biggr),
\nn\\[.5em]
\lefteqn{ \ri\partial\SZ + \ri\MZ\Schi 
   - \sum_\pm (\pm e)\frac{\cw}{\sw} \dzd{\SWpm^\mu}{\SWpm^\mu} 
   - \sum_\pm (\pm e)\frac{\cw^2-\sw^2}{2\cw\sw} \dzd{\Sphipm}{\Sphipm} }
\quad \nl &&{}
+ \frac{\ri e}{2\cw\sw} \biggl( \dzd{\SH}{\Schi} - \dzd{\Schi}{\SH} \biggr) 
   + e\sum_f \biggl( \dzd{\Sf}(v_f+a_f\ga_5)\Sf + 
		     \Sfbar(v_f-a_f\ga_5)\dzd{\Sfbar} \biggr)
\nl &=&
- \frac{\ri}{\xiZhat}(\quabla+\xiZhat\MZ^2)
   \biggl( \partial^\mu\dzd{\SZ^\mu} - \xiZhat\MZ\dzd{\Schi} \biggr)  
- \frac{\ri e\MZ}{2\cw\sw}\dzd{\SH}
   \biggl( \partial^\mu\dzd{\SZ^\mu} - \xiZhat\MZ\dzd{\Schi} \biggr)
\nl &&{}
- \sum_\pm \frac{(\pm e)}{\xiWhat}
   \biggl( \frac{\cw}{\sw}\dzd{\SWmp^\mu}\partial^\mu
   \pm \ri\xiWhat\MW\frac{\cw^2-\sw^2}{2\cw\sw}\dzd{\Sphimp} \biggr)
   \biggl( \partial^\nu\dzd{\SWpm^\nu} 
   \pm \ri\xiWhat\MW\dzd{\Sphipm^\mu} \biggr),
\nn\\[.5em]
\lefteqn{ \ri\partial\SWmp \pm \MW{\Sphimp}
   \mp e\dzd{\SWpm^\mu}\biggl( {\SA^\mu}-\frac{\cw}{\sw}{\SZ^\mu} \biggr)
   \pm e\biggl( \dzd{\SA^\mu}-\frac{\cw}{\sw}\dzd{\SZ^\mu} \biggr){\SWmp^\mu} }
\quad \nl && {}
\mp \frac{e}{2\sw}\dzd{\Sphipm} ({\SH} \pm \ri{\Schi})
   \pm \frac{e}{2\sw}\biggl( \dzd{\SH} \pm \ri\dzd{\Schi} \biggr) \Sphimp 
\nl &&{}
+ \frac{e}{\sqrt{2}\sw} \sum_{(f_+,f_-)} 
   \biggl( \dzd{\Sfpm}\frac{1 +\ga_5}{2}\Sfmp
           + \Sfbarpm\frac{1-\ga_5}{2}\dzd{\Sfbarmp} \biggr)
\nl &=&
- \frac{\ri}{\xiWhat}(\quabla+\xiWhat\MW^2)
   \biggl( \partial^\mu\dzd{\SWpm^\mu}
   \pm \ri\xiWhat\MW\dzd{\Sphipm} \biggr) 
\quad \nl &&{}
\mp \frac{e}{\xiWhat}\biggl[
   \biggl( \dzd{\SA^\mu} - \frac{\cw}{\sw}\dzd{\SZ^\mu} \biggr)\partial^\mu
   \pm \ri\xiWhat\frac{\MW}{2\sw}\biggl( \dzd{\SH}\pm\ri\dzd{\Schi} \biggr)
   \biggr] 
\nl &&{} \qquad\times
\biggl( \partial^\nu\dzd{\SWpm^\nu} \pm \ri\xiWhat\MW\dzd{\Sphipm} \biggr)
   \pm \frac{e}{\xiAhat}\dzd{\SWpm^\mu}\partial^\mu\partial^\nu\dzd{\SA^\nu} 
\nl &&{}
\mp \frac{e}{\xiZhat}\biggl( \frac{\cw}{\sw}\dzd{\SWpm^\mu}\partial^\mu
   \mp \ri\xiZhat\frac{\MZ}{2\sw}\dzd{\Sphipm} \biggr)
   \biggl( \partial^\nu\dzd{\SZ^\nu} - \xiZhat\MZ\dzd{\Schi} \biggr), 
\eeqar
where $\vf = (I^3_{\rw,f} - 2\sw^2 Q_f)/(2\sw\cw)$
and $\af = I^3_{\rw,f}/(2\sw\cw)$ are the vector and axial-vector
couplings of the \PZ~boson to the fermion $f$ with relative charge $Q_f$ and 
third component of weak iso-spin $I^3_{\rw,f}$.
In \refeq{eq:WIZ} $f_\pm$ denote the 
fermionic iso-spin partners with 
iso-spins $\pm1/2$,
and the sum over $(f_+,f_-)$ runs over all iso-spin doublets.

By taking functional derivatives of 
\refeq{eq:WIZ} with respect to the
sources one obtains Ward identities for connected Green functions.
These Ward identities hold for any fixed loop order in perturbation theory
exactly. 
This is evident if one expands everything 
including propagators in powers of the coupling constant $e$.
However, {\em the Ward identities hold as well after Dyson summation of
the self-energy corrections 
if the inverse propagators, which are just the two-point vertex functions, 
are calculated in the same loop order as all other vertex functions.} 
In order to see this, one has to go back to the background-field effective 
action $\Ga$ and its Ward identities \refeq{eq:WIGamma}. 
As these are linear in $\Ga$, its $n$-loop approximation 
$\Ga|_{\mathrm{n-loop}}$ fulfills exactly the same Ward identities. 
Substituting
$\Ga|_{\mathrm{n-loop}}$ instead of $\Ga$ into \refeq{eq:Gammafull} 
analogously defines $\Ga^\full|_{\mathrm{n-loop}}$, which in turn defines
$\Zc|_{\mathrm{n-loop}}$ via a Legendre transformation, as written down
for $\Zc$ in \refeq{eq:Zc}--\refeq{eq:ZtoF}. 
Consequently,
$\Zc|_{\mathrm{n-loop}}$ is the generating functional for connected Green 
functions built of vertex functions in $n$-loop approximation 
and propagators that are defined as the inverse two-point vertex
functions in the same approximation, \ie all propagators include the 
Dyson-resummed \ses\ in $n$-loop approximation. By construction all
previous relations remain valid if $\Ga$, $\Ga^\full$, and $\Zc$ are
replaced by $\Ga|_{\mathrm{n-loop}}$, $\Ga^\full|_{\mathrm{n-loop}}$, and
$\Zc|_{\mathrm{n-loop}}$, respectively. 
This proves,
in particular, that $\Zc|_{\mathrm{n-loop}}$, which
involves Dyson summation, fulfills the Ward identities \refeq{eq:WIZ} 
exactly.

Consequently, Dyson summation within the BFM does not disturb the high-energy
behavior of physical amplitudes; 
in particular, gauge cancelations are
not violated. This feature is not present in the conventional formalism.
We note that the BFM vertex functions still depend on 
the quantum gauge parameter $\xi_Q$.
However, the logarithmic contributions to
the self-energies that dominate at high energies are gauge-independent
and universal \cite{bgflett}; they are in fact governed by the renormalization 
group.

The previous considerations show that the BFM allows Dyson summation and
thus, in particular, the introduction
of finite-width effects without spoiling the Ward identities.
Unfortunately, a dependence on the quantum gauge 
parameter remains,
which
cannot be fixed on physical grounds. 
So far---to the 
best of our knowledge---there is no 
prescription available that yields a unique 
unambiguous answer in the general case. 
However, for particles that decay only into fermions,
such as the Z and W~bosons,
a practical way consists in resumming only the fermionic one-loop
corrections \cite{bhf1}. Since these
are identical in the BFM and the conventional approach, and the complete
one-loop corrections are just the sum of the fermionic and bosonic
corrections, our analysis provides an independent proof of the
fermion-loop scheme for the treatment of finite width-effects
in tree-level amplitudes.

\subsection{Propagators}

In the following, we need the explicit form of the
Ward identities for the two-point functions,
\ie the propagators, of the gauge and scalar bosons.
These are obtained by differentiating \refeq{eq:WIZ} with respect to 
the corresponding sources, putting all sources to 
zero and using $\de\Zc/\de J_{\Fhat}|_{J_{\Fhat}=0}=0$. 
Whereas this last equation is clear for all other fields it is enforced
for the physical Higgs field 
by a renormalization condition (vanishing tadpole).
Introducing a $4\times4$ matrix 
\beqar
G^{0}_{(\mu\nu)} = \left( \barr{cccc}
G^{\Ahat\Ahat}_{\mu\nu} & G^{\Ahat\Zhat}_{\mu\nu} & 
G^{\Ahat\chihat}_{\mu}  & G^{\Ahat\Hhat}_{\mu} \\
G^{\Zhat\Ahat}_{\mu\nu} & G^{\Zhat\Zhat}_{\mu\nu} & 
G^{\Zhat\chihat}_{\mu}  & G^{\Zhat\Hhat}_{\mu} \\
G^{\chihat\Ahat}_{\nu}  & G^{\chihat\Zhat}_{\nu} & 
G^{\chihat\chihat}      & G^{\chihat\Hhat} \\
G^{\Hhat\Ahat}_{\nu}    & G^{\Hhat\Zhat}_{\nu} & 
G^{\Hhat\chihat}        & G^{\Hhat\Hhat} 
\earr\right)
\eeqar
for the neutral boson propagators
and a $2\times2$ matrix 
\beq
G^\pm_{(\mu\nu)} = \left( \barr{cc}
G^{\What^\pm\What^\mp}_{\mu\nu} & G^{\What^\pm\phihat^\mp}_{\mu}  \\
G^{\phihat^\pm\What^\mp}_{\nu} & G^{\phihat^\pm\phihat^\mp}      \\
\earr\right)
\eeq
for the charged boson propagators,
the Ward identities can be compactly written as
\beqar\label{eq:WIprop}
(k^\mu,0,0,0)G^{0}_{(\mu\nu)} &=& -\ri\xiAhat\frac{1}{k^2}(k_\nu,0,0,0), \nl
(0,k^\mu,\ri\xiZhat\MZ,0) G^{0}_{(\mu\nu)} &=& 
-\ri\xiZhat\frac{1}{k^2-\xiZhat\MZ^2}(0,k_\nu,-\ri\MZ,0),\nl
(k^\mu,\pm\xiWhat\MW) G^\pm_{(\mu\nu)} &=& 
-\ri\xiWhat\frac{1}{k^2-\xiWhat\MW^2}(k_\nu,\mp\MW),
\eeqar
where we have turned to momentum space for later convenience.
In \refeq{eq:WIprop} $k$ is the momentum flowing through the two-point 
functions. Our conventions for the Fourier transformation from
coordinate to momentum space are summarized in \refapp{app:momconv}.

Note that in the conventional formalism the left-hand sides of these
relations are much more complicated and involve ghost contributions
\cite{Ba90,He92}.

\subsection{Amputated connected Green functions}
\label{sse:ampGF}

The Ward identities \refeq{eq:WIZ} involve four different types of terms:
The effective action gives rise to terms involving
$J$ or $J\de \Zc/\de J$, the gauge-fixing term introduces terms
containing $\de \Zc/\de J$ or $(\de \Zc/\de J)^2$.
The $J$ terms obviously drop out when more than one functional
derivative is taken, i.e.\ they merely contribute to the Ward identities
for two-point functions, which have been given in the previous section. 
In \refapp{app:WIproof} we proof that
the $J\de\Zc/\de J$ and $(\de\Zc/\de J)^2$ terms do not contribute 
to Green functions after amputating and putting all external
physical fields on their mass shell. 
Consequently, the Ward identities \refeq{eq:WIZ} imply
\beqar\label{eq:WIZonsmom}
0 &=& k^\mu\dzd{\SA^\mu} + \ \OVT \,, \nl
0 &=& \biggl( k^\mu\dzd{\SZ^\mu}+\ri\xiZhat\MZ\dzd{\Schi} \biggr)
    + \ \OVT \,, \nl
0 &=& \biggl( k^\mu\dzd{\SWpm^\mu}\pm\xiWhat\MW \dzd{\Sphipm} \biggl)
    {}+{} \ \OVT \,, 
\eeqar
where $\OVT$ (on-shell vanishing terms) 
stands for terms that vanish after taking derivatives with
respect to physical fields and subsequent amputation and on-shell projection.

Equations 
\refeq{eq:WIZonsmom} represent the
identity \refeq{WIET} for one gauge-fixing condition.
The generalization to more external ``gauge-fixing legs'' is also shown
in \refapp{app:WIproof}.

Note that the Ward identities \refeq{eq:WIZonsmom} and their
generalizations 
are identical to the identities \refeq{WIET} of the conventional formalism. 
This is due to the fact that
they hold for on-shell physical fields and that the gauge-fixing term
for the background fields is identical to the one in the conventional
formalism.

In order to derive the ET from 
\refeq{eq:WIZonsmom}, we have to amputate 
the external legs that correspond to the 
gauge-fixing operators.
Because of the mixing 
between
longitudinal gauge bosons and 
would-be Goldstone bosons, this amputation must be done carefully.
Marking amputated external legs by lowering  the corresponding field
index of the Green function, we can write the relation between 
non-amputated and amputated vertex functions for neutral bosons as follows
\beq\label{eq:ampdef}
\left( \barr{c} G^{\Ahat\ldots}_{\ldots,\mu} \\ G^{\Zhat\ldots}_{\ldots,\mu} \\ 
G^{\chihat\ldots}_{\ldots} \\ G^{\Hhat\ldots}_{\ldots}\earr \right) 
= G^{0}_{(\mu\nu)} 
\left(\barr{c}G^{\ldots,\nu}_{\Ahat\ldots} \\ G^{\ldots,\nu}_{\Zhat\ldots} 
\\ G^{\ldots}_{\chihat\ldots} \\ G^{\ldots}_{\Hhat\ldots}\earr\right), \qquad
\left( \barr{c} G^{\What^\pm\ldots}_{\ldots\mu} \\ 
       G^{\phihat^\pm\ldots}_{\ldots}
\earr\right) = G_{(\mu\nu)}^\pm 
\left(\barr{c} 
G^{\ldots,\nu}_{\What^\pm\ldots} \\ G^{\ldots}_{\phihat^\pm\ldots} 
\earr\right),
\eeq
where the dots 
indicate the remaining amputated and non-amputated external legs.

In this matrix notation the Ward identities 
\refeq{eq:WIZonsmom} 
read
\beqar\label{eq:WIGmat}
&& (k^\mu,0,0,0)
\left( \barr{c} G^{\Ahat}_{\ldots,\mu} \\
G^{\Zhat}_{\ldots,\mu} \\
G^{\chihat}_{\ldots} \\
G^{\Hhat}_{\ldots}\earr \right) = 0,\qquad
(0,k^\mu,\ri\hat\xi_Z\MZ,0)
\left( \barr{c} G^{\Ahat}_{\ldots,\mu} \\
G^{\Zhat}_{\ldots,\mu} \\
G^{\chihat}_{\ldots} \\
G^{\Hhat}_{\ldots}\earr \right) = 0, \  \nl[3ex]
&& (0,k^\mu,\pm\hat\xi_W\MW,0)
\left( \barr{c} G^{\What^\pm}_{\ldots,\mu}
\\ G^{\phihat^\pm}_{\ldots} \earr\right) = 0, 
\eeqar
assuming that all physical external legs are already amputated and put on-shell.
Upon inserting \refeq{eq:ampdef} into \refeq{eq:WIGmat} 
and using \refeq{eq:WIprop}, 
we obtain the Ward identities for on-shell amputated Green functions,
\beq\label{eq:WIET}
k^\nu G_{\Ahat\ldots,\nu} = 0, \qquad
k^\nu G_{\Zhat\ldots,\nu} = \ri\MZ G_{\chihat\ldots}, \qquad
k^\nu G_{\What^\pm\ldots,\nu} = \pm\MW G_{\phihat^\pm\ldots}.
\eeq
The corresponding Ward identities in the conventional formalism involve
extra factors, which depend on renormalization constants and unphysical
gauge-boson and 
would-be Goldstone-boson self-energies. These factors can be eliminated 
by suitably tuning the renormalization in the unphysical sector 
\cite{Ba90,He92}. In the BFM these factors are naturally absent owing to the 
background-field gauge invarinace.

The first of the Ward identities \refeq{eq:WIET} expresses
transversality for on-shell photons, the other two imply the ET for the
massive gauge bosons, as will be described in \refse{se:et}.

The Ward identities for 
arbitrarily many gauge-fixing legs follow from
\refeq{eq:multWI} and \refeq{eq:ampdef} exactly in the same way.

\section{WAVE-FUNCTION RENORMALIZATION}
\label{se:physsta}

As already mentioned, the background fields are assumed
to be renormalized as described in \citere{bgflong}. This 
choice implies that the field renormalization constants are adjusted to
the parameter renormalization constants such that unrenormalized and
renormalized Ward identities are formally identical. However, 
it also implies that---except for the photon---the residues of the 
propagators differ from one, and that
the (on-shell) \PZ-boson field mixes with 
the photon field at $k^2=\MZ^2$. Consequently, we have to carry out a
(UV-finite) wave-function renormalization when constructing $S$-matrix 
elements from on-shell amputated Green functions.%
\footnote{The necessity of a (UV-finite) wave-function renormalization
in addition to a field renormalization which removes the UV divergences
is not a peculiarity of the BFM. 
It arises  for instance also in the minimal renormalization scheme of the SM,
where multiplets of fields are renormalized by a single field renormalization 
constant \cite{bhs}.
}

In the charged sector the situation is relatively simple, since the 
physical components of the W-boson field $\What^\pm$ do
not mix with any
other field. The mixing with the fields $\phihat^\pm$ only takes place
for the unphysical components of $\What^\pm$.
An $S$-matrix element involving an external $\PW$ boson
can only differ in normalization from the corresponding on-shell
amputated Green function which is contracted with the polarization vector
$\veps_{\PW}(k)$ of the external \PW~boson,
\beq\label{eq:SMECC}
\langle\ldots|S|W^\pm(k)\ldots\rangle =
\RW\veps_{\PW,\mu}(k) G_{\What^\pm\ldots}^\mu(k,\ldots).
\eeq
The wave-function renormalization constant $\RWW$ is fixed by requiring
that the pole of the transverse part of the \PW-boson propagator
has residue one, or equivalently
\beq\label{eq:RWcond}
\RWW \Re\left.\left\{\frac{\ri\Ga^{\What^+\What^-}_{\mu\nu}(k)}
{k^2-\MW^2}\right\} \veps^\nu_{\PW}(k)
\right|_{k^2=\MW^2} = \veps_{\PW,\mu}(k).
\eeq
Using the decomposition of two-point functions into transverse and 
longitudinal parts,
\beq\label{eq:2ptdecomp}
\Ga^{\Vhat\Vhat'}_{\mu\nu}(k) = 
\left(g_{\mu\nu}-\frac{k_\mu k_\nu}{k^2}\right) \Ga^{\Vhat\Vhat'}_\rT(k^2)
+\frac{k_\mu k_\nu}{k^2} \Ga^{\Vhat\Vhat'}_\rL(k^2),
\eeq
the condition \refeq{eq:RWcond} for $\RWW$ implies
\beq\label{eq:RW}
\RWW^{-1} 
= \left.\Re\left\{\frac{\partial}{\partial k^2}
\ri\Ga_\rT^{\What^+\What^-}(k^2)\right\}\right|_{k^2=\MW^2}
= \Re\{\ri\Ga_\rT^{\prime\What^+\What^-}(\MW^2)\}.
\eeq
We remind the reader that all quantities are renormalized and that,
in particular, the poles of the propagators are at the physical masses.

In the neutral sector things are complicated by the 
mixing between the photon and the $\PZ$ boson. 
The mixing with the scalar fields $\Hhat$, $\chihat$ again only takes
place in the unphysical degrees of freedom and need not 
to be considered.
The $S$-matrix elements involving a photon or a \PZ~boson 
result from superpositions 
of the corresponding amputated Green functions,
\beqar\label{eq:SMENC}
\langle\ldots|S|A(k)\ldots\rangle &=&
\veps_{\PA,\mu}(k)\Bigl[R_{\Ahat\Ahat}^{1/2} G_{\Ahat\ldots}^\mu(k,\ldots)
+ R_{\Zhat\Ahat}^{1/2} G_{\Zhat\ldots}^\mu(k,\ldots)\Bigr], \nn\\ 
\langle\ldots|S|Z(k)\ldots\rangle &=&
\veps_{\PZ,\mu}(k)\Bigl[R_{\Ahat\Zhat}^{1/2} G_{\Ahat\ldots}^\mu(k,\ldots)
+ R_{\Zhat\Zhat}^{1/2} G_{\Zhat\ldots}^\mu(k,\ldots)\Bigr]. 
\eeqar
The generalization to more gauge bosons is obvious.
The wave-function renormalization constants $R_{\Vhat\Vhat'}$ are fixed by 
requiring that 
one-particle states are normalized and propagate without mixing with other 
fields. More explicitly, this means that the matrix propagator for the
photon and Z-boson fields is diagonal at $k^2=0$ and $k^2=\MZ^2$, and the
corresponding residues at the propagator poles are equal to one.
For the amputated Green functions or vertex functions these conditions
read
\beqar\label{eq:RAZcond}
\veps_{\PA,\mu}(k) &=& 
\Re\left.\left\{ \frac{\ri}{k^2}\left[
R_{\Ahat\Ahat}\Ga^{\Ahat\Ahat}_{\mu\nu}(k)
+2R_{\Ahat\Ahat}^{1/2}R_{\Zhat\Ahat}^{1/2}\Ga^{\Zhat\Ahat}_{\mu\nu}(k)
+R_{\Zhat\Ahat}\Ga^{\Zhat\Zhat}_{\mu\nu}(k)
\right] \right\} \veps^\nu_{\PA}(k) \right|_{k^2=0},
\nn\\[.5em] 
0 &=&
\Re\left.\left\{ \ri \left[
R_{\Ahat\Ahat}^{1/2}R_{\Ahat\Zhat}^{1/2}\Ga^{\Ahat\Ahat}_{\mu\nu}(k)
+\left(R_{\Ahat\Ahat}^{1/2}R_{\Zhat\Zhat}^{1/2}+
  R_{\Ahat\Zhat}^{1/2}R_{\Zhat\Ahat}^{1/2}\right)\Ga^{\Zhat\Ahat}_{\mu\nu}(k)
\right.\right.\right. \nn\\[.3em] && \left.\left.\left.\hspace{17em}
+R_{\Zhat\Zhat}^{1/2}R_{\Zhat\Ahat}^{1/2}\Ga^{\Zhat\Zhat}_{\mu\nu}(k)
\right] \right\}
\veps^\nu_{\PA,\PZ}(k)
\right|_{k^2=0,\MZ^2},
\nn\\[.5em]
\veps_{\PZ,\mu}(k) &=& 
\Re\left.\left\{ \frac{\ri}{k^2-\MZ^2}\left[
R_{\Ahat\Zhat}\Ga^{\Ahat\Ahat}_{\mu\nu}(k)
+2R_{\Ahat\Zhat}^{1/2}R_{\Zhat\Zhat}^{1/2}\Ga^{\Zhat\Ahat}_{\mu\nu}(k)
+R_{\Zhat\Zhat}\Ga^{\Zhat\Zhat}_{\mu\nu}(k)
\right] \right\} \veps^\nu_{\PZ}(k) \right|_{k^2=\MZ^2}.
\nn\\
\eeqar
Inserting the decomposition \refeq{eq:2ptdecomp} for the two-point
functions into \refeq{eq:RAZcond}, the constants $R_{\Vhat\Vhat'}$ can
be expressed in terms of 
the transverse parts $\Ga^{\Vhat\Vhat'}_{\rT}$.
Using in addition the equations
\beq \label{eq:WIA}
\Ga^{\Ahat\Ahat}_\rT(0) = 
\Ga^{\Ahat\Zhat}_\rT(0) = 0,
\qquad\Ga^{\prime\Ahat\Ahat}_\rT(0) = -\ri,
\eeq
which follow from the Ward identities and the on-shell renormalization
condition for the electric charge \cite{bgflong},
we find 
\beqar\label{eq:Rnc}
R_{\Ahat\Ahat}^{1/2} = 1, & \qquad &
R_{\Ahat\Zhat}^{1/2} = -\frac{\Re\{\ri\Ga_\rT^{\Ahat\Zhat}(\MZ^2)\}}
{\Re\{\ri\Ga_\rT^{\Ahat\Ahat}(\MZ^2)\}} R_{\Zhat\Zhat}^{1/2}, \nl
R_{\Zhat\Ahat}^{1/2} = 0, & \qquad &
R_{\Zhat\Zhat}^{-1}  = \Re\{\ri\Ga_\rT^{\prime\Zhat\Zhat}(\MZ^2)\}
-2\Re\{\ri\Ga_\rT^{\prime\Ahat\Zhat}(\MZ^2)\}
\frac{\Re\{\ri\Ga_\rT^{\Ahat\Zhat}(\MZ^2)\}}
{\Re\{\ri\Ga_\rT^{\Ahat\Ahat}(\MZ^2)\}}
\nl && \qquad\qquad
{}+\Re\{\ri\Ga_\rT^{\prime\Ahat\Ahat}(\MZ^2)\}
\left(\frac{\Re\{\ri\Ga_\rT^{\Ahat\Zhat}(\MZ^2)\}}
{\Re\{\ri\Ga_\rT^{\Ahat\Ahat}(\MZ^2)\}}\right)^{\!\!2} .
\eeqar
This shows, in particular, that, as a consequence of the Ward identities
\refeq{eq:WIA}, the renormalized photon has residue one
and does not mix with the \PZ~boson for $k^2=0$.

We note in passing that the \PZ-boson mass is fixed by the condition
\beq
0 = \Re\left\{ \ri\Ga^{\Zhat\Zhat}_\rT(\MZ^2) 
- \ri\left(\Ga^{\Ahat\Zhat}_\rT(\MZ^2)\right)^2
/\Ga^{\Ahat\Ahat}_\rT(\MZ^2)\right\},
\eeq
which is invariant under the transformation 
related to the wave-function renormalization.

Decomposing the transverse parts of the two-point functions into 
lowest-order contributions and transverse self-energies,
\beq
\Ga^{\Vhat\Vhat'}_{\rT}(k^2) = 
-\ri(k^2-\MV^2)\de_{\Vhat\Vhat'} -\ri \Si^{\Vhat\Vhat'}_{\rT}(k^2) ,
\eeq
yields simple one-loop expressions for the $R$ factors for external
W and Z~bosons,
\beqar\label{eq:Ronel}
\RW &=& 
  1-\frac{1}{2}\Re\left\{\Si^{\prime\What^+\What^-}_\rT(\MW^2)\right\} 
  +{\cal O}(\alpha^2),\nl
R_{\Zhat\Zhat}^{1/2} &=& 
  1-\frac{1}{2}\Re\left\{\Si_\rT^{\prime\Zhat\Zhat}(\MZ^2)\right\}
  +{\cal O}(\alpha^2),\nl
R_{\Ahat\Zhat}^{1/2} &=& 
  -\frac{1}{\MZ^2}\Re\left\{\Si_\rT^{\Ahat\Zhat}(\MZ^2)\right\}
  +{\cal O}(\alpha^2).
\eeqar
We recall that BFM vertex functions, and thus also the $R$ factors, depend 
on the quantum gauge parameter $\xi_Q$, which enters by fixing the gauge of 
the quantum fields.
Of course, $\xi_Q$ cancels in any complete
loop order when calculating $S$-matrix elements.
The explicit one-loop results for the self-energies needed for the $R$
factors according to \refeq{eq:Ronel} are given in \refapp{app:nomfac}
in 't~Hooft--Feynman gauge ($\xi_Q=1$).
Finally, we note that $\RW$
possesses an IR-divergent contribution in analogy to the corresponding
wave-function renormalization constant 
in the conventional formalism
\cite{bhs,Dehab,Ao82}. 

\section{THE EQUIVALENCE THEOREM }
\label{se:et}

We have now all ingredients to derive the ET.
The longitudinal polarization vector of a massive gauge boson 
($V_a=W,Z$) with momentum $k$ can be decomposed as
\beq
\veps_{a,\rL}^\mu(k) = \frac{k^\mu}{M_a} + v_a^\mu(k),  \qquad
v_a^\mu(k) = \O\left(\frac{M_a}{k^0}\right),
\eeq
i.e.\ its leading part at high energies is proportional to the momentum.
Inserting this decomposition into 
the expressions \refeq{eq:SMECC} and \refeq{eq:SMENC} for the 
$S$-matrix elements, the Ward identities \refeq{eq:WIET} 
directly imply 
for one external longitudinal gauge boson
\beqar \label{eq:ET0}
\langle\ldots|S|W_\rL^\pm(k)\ldots\rangle &=&
\pm R_{\What}^{1/2} G_{\phihat^\pm\ldots} +
v_W^\mu(k)R_{\What}^{1/2}G_{\What^\pm\ldots} ,\nl
\langle\ldots|S|Z_\rL(k)\ldots\rangle &=&
\ri R_{\Zhat\Zhat}^{1/2} G_{\chihat\ldots} +
v_Z^\mu(k)R_{\Zhat\Zhat}^{1/2}G_{\Zhat\ldots} +
v_Z^\mu(k)R_{\Ahat\Zhat}^{1/2}G_{\Ahat\ldots}.
\eeqar
Since unitarity ensures that $S$-matrix elements in the SM do not
grow with powers of the gauge-boson energy $k^0$ in the high-energy limit,
the contributions of $v_a^\mu(k)$ are of order $\O(M_a/k^0)$ 
and thus negligible,
\beqar\label{eq:ET}
\langle\ldots|S|W_\rL^\pm(k)\ldots\rangle &=&
\pm R_{\What}^{1/2} G_{\phihat^\pm\ldots} +
\O\left(\frac{\MW}{k^0}\right),\nl
\langle\ldots|S|Z_\rL(k)\ldots\rangle &=&
\ri R_{\Zhat\Zhat}^{1/2} G_{\chihat\ldots} +
\O\left(\frac{\MZ}{k^0}\right).
\eeqar
Equations \refeq{eq:ET} represent the precise formulation of the ET for
the SM within the framework of the BFM.

The case of more longitudinal gauge bosons can be treated in the same 
way as in \citere{Ch85}, the only difference being the factors $R$.
As is easily seen, for each external longitudinal $\PW^\pm$ boson 
an extra factor $\smash{R_{\What}^{1/2}}$ and for each 
external longitudinal \PZ~boson an extra factor $R_{\Zhat\Zhat}^{1/2}$
has to be introduced. This concludes the derivation of the ET 
for the SM in the BFM.

The form \refeq{eq:ET} of the ET clearly displays 
one of the advantages of its
formulation within the framework of the BFM. The correction factors,
which modify the na\"\i ve form of the ET, are simply given by the
residues of the renormalized
massive gauge-boson propagators, which are needed for
the calculation of $S$-matrix elements anyhow. 

As pointed out in \refse{se:wi},
the underlying Ward identities \refeq{eq:WIZ} and
\refeq{eq:WIET} are not only valid order by order in perturbation theory 
but also after Dyson resummation.
Consequently, also {\em the ET \refeq{eq:ET} within the BFM is 
valid after Dyson summation}.

The contributions of $v^\mu$ vanish
owing to unitarity only if the energy $E$ 
of the physical process is large compared to
all masses present in the SM including the Higgs-boson mass, $\MH$. 
However, it is often very interesting to know to which order in $E^n\MH^m$ 
the ET \refeq{eq:ET} is still valid also for large $\MH$ 
or how it has to be modified in this case. This can be determined by
a power-counting method developed in \citeres{Ve90,Gr95b}.
Although this method was worked out for the conventional formalism, it
is also applicable in the BFM since the leading powers in propagators
and couplings are identical in both approaches.

The explicit expressions for the factors $R$ in the SM in the BFM at
one-loop order contain no terms of order $\MH^2/\MW^2$
for a large Higgs-boson mass. 
As a consequence these factors can be put to one in this
limit if one is only interested in terms 
which are enhanced by powers of $E^2/\MW^2$ or $\MH^2/\MW^2$.
For $\MH\gg\MW$ the factors $R$ get $\log\MH$ corrections, which are
explicitly given in \refapp{app:nomfac}.

\section{NON-LINEARLY REALIZED HIGGS SECTOR OF THE STANDARD MODEL}
\label{se:nlscalar}

In the previous sections we have dealt with the commonly used {\it
linear} realization of the Higgs sector of the SM, 
where the scalar Higgs doublet $\Phi$ is represented as 
\beq
\Phi=\frac{1}{\sqrt{2}}\left((v+H){\bf 1}+\ri\phi^a\tau^a\right), \qquad
\phi^\pm = \frac{1}{\sqrt{2}}\left(\phi^2\pm\ri\phi^1\right), \qquad
\chi = -\phi^3.
\label{eq:philin}
\eeq
In \refeq{eq:philin} we have adopted the matrix notation of \citere{sdcgk}
with 
$\tau^a$ denoting the Pauli matrices.
In the linear representation the physical Higgs field $H$ is not
gauge-invariant. Alternatively, the scalar field $\Phi$ can be {\it
non-linearly} represented \cite{nlhiggs} as
\beq\label{eq:phinl}
\Phi=\frac{1}{\sqrt{2}}(v+H)U \qquad \mbox{with} \quad
U=\exp\left(\frac{\ri}{v}\phi^a\tau^a\right),
\eeq
where the would-be Goldstone-boson fields $\phi^a$ parameterize the unitary 
matrix $U$. The explicit parameterization of $U$ is not uniquely determined
but the above exponential form is very convenient.
The non-linear realization has the interesting property
that the physical Higgs field $H$ is gauge-invariant and
that the scalar 
self-couplings 
do not involve the unphysical would-be Goldstone-boson fields but only $H$.

The application of the BFM to the non-linear 
realization of the Higgs sector
(together
with the corresponding gauge-invariant renormalization) was worked out in
\citere{sdcgk} and also briefly discussed in \citere{bgfring}. 
In the BFM approach the main difference between linear and non-linear 
realization relies in the splitting of the would-be Goldstone-boson
fields into background and quantum parts: the unitary matrix $U$ is split
multiplicatively ($U\to\hat UU$).
Using the
above treatment of the linear realization as guideline, the ET for the
non-linear realization can be derived exactly in the same way. The
actual calculation degenerates to a straightforward exercise so that it
suffices to briefly describe the single steps.

The starting point is the derivation of the Ward identities, which
follow from the invariance of the BFM effective action under background
gauge transformations, as expressed by \refeq{eq:WIGamma}. As far as the
Ward identities are concerned, the only difference between linear and
non-linear realization lies in the explicit form of the 
gauge transformations of the background scalar fields $\Hhat$,
$\phihat^\pm$, $\chihat$. For the linear realization \refeq{eq:philin}
they are explicitly given by Eq.~(21) of \citere{bgflong}, for the
non-linear one \refeq{eq:phinl} they can be deduced
from \citere{sdcgk}:
\beqar\label{eq:nlgtrafo}
\de\Hhat &=& 0, \nn\\
\de\phihat^\pm &=& \pm\ri\MW\de\thetahat^\pm
-\frac{e}{2\sw}\chihat\de\thetahat^\pm
\mp\ri e \phihat^\pm \left(\de\thetahat^A
   -\frac{\cw^2-\sw^2}{2\cw\sw}\de\thetahat^Z\right) \nn\\
&&{} + g\left(\frac{\phihat^a\phihat^a}{v^2}\right) 
   \frac{e^2}{4\sw^2\MW} \left[ 
   \pm\ri\phihat^\pm\phihat^\pm\de\thetahat^\mp
   \pm\ri\phihat^+\phihat^-\de\thetahat^\pm
   \pm\ri\chihat^2\de\thetahat^\pm
   +\frac{1}{\cw}\phihat^\pm\chihat\de\thetahat^Z \right], \nn\\[.5em]
\de\chihat &=& -\MZ\de\thetahat^Z
+\frac{e}{2\sw}(\phihat^+\de\thetahat^-+\phihat^-\de\thetahat^+) \nn\\
&&{} + g\left(\frac{\phihat^a\phihat^a}{v^2}\right) 
   \frac{e^2}{4\sw^2\MW} \left[
   -\ri\chihat(\phihat^-\de\thetahat^+-\phihat^+\de\thetahat^-)
   -\frac{2}{\cw}\phihat^+\phihat^-\de\thetahat^Z \right].
\eeqar
Owing to the non-linearity of $\hat\Phi$, the gauge variations 
$\de\phihat^\pm$ and $\de\chihat$ involve arbitrary powers of $\phihat^\pm$
and $\chihat$ occurring in the function
\beq
g(x) = \frac{\cot(\sqrt{x})}{\sqrt{x}} - \frac{1}{x} =
-\frac{1}{3}-\frac{x}{45}+\cdots \quad \mbox{with} \quad
x=\frac{\phihat^a\phihat^a}{v^2}=
\frac{e^2}{4\sw^2\MW^2}(2\phihat^+\phihat^-+\chihat^2).
\eeq
The $g$-independent terms in \refeq{eq:nlgtrafo} 
are equal to the gauge variations in the linear
representation with the physical Higgs field $\Hhat$ omitted.
The procedure of fixing the gauge of the background fields and
performing the Legendre transformation to the generating functional
$\Zc$ of connected Green functions is carried out as above; in particular,
\refeq{eq:WIGamma} -- \refeq{eq:WIGammafull} and \refeq{eq:Zc}
-- \refeq{eq:ZtoF} apply literally. In analogy to the derivation of
\refeq{eq:WIZ} one obtains the functional form of the Ward
identities for connected Green functions, which 
in contrast to \refeq{eq:WIZ} involves arbitrary
powers of $(\de\Zc/\de J_{\phihat^\pm})$ and $(\de\Zc/\de J_{\chihat})$.
However, the Ward identities \refeq{eq:WIprop} for the (renormalized) 
two-point functions are identical in both realizations. Moreover, the
identities \refeq{eq:WIZonsmom} for on-shell Green functions 
and their generalization for more gauge-fixing legs
follow by the same reasoning as described in 
\refapp{app:WIproof} implying \refeq{eq:WIET} for on-shell amputated
Green functions. Obviously, the 
wave-function renormalization and the construction of the $S$ matrix, as
described in \refse{se:physsta},
do not rely on a specific property of the scalar sector. In summary, 
we arrive again at the ET in the form of \refeq{eq:ET}.

It is quite easy to see that the explicit representation of the Higgs
field $\Phi$ and its behavior under background gauge transformations in
general are not important for the basic Ward identities 
\refeq{eq:WIZonsmom}--\refeq{eq:WIET}. 
The only relevant terms in the gauge transformation of the would-be
Goldstone-boson
fields are the constant contributions, i.e.\ the ones which do
not depend on the fields $\phi^a$.
Nevertheless, we have given the
explicit form \refeq{eq:nlgtrafo} of the background gauge transformations 
for the scalar fields corresponding to the non-linear realization 
\refeq{eq:phinl} in order to illustrate some interesting features.
Comparing linear and non-linear realizations, one can see that all Ward
identities that depend on the gauge-fixing term of the photon
are identical in both
cases, as the variations of the scalar fields with $\de\thetahat^A$
coincide. If gauge-fixing terms of the massive gauge fields are
involved, the Ward identities in general are different. Moreover, the
gauge invariance of $\Hhat$ implies that neither $\SH$ nor
$\de\Zc/\de\SH$ occur in the functional form [the analogue of
\refeq{eq:WIZ}] of the Ward identities. This means that all external
Higgs-boson legs result from derivatives with respect 
to physical particles and occur in the same way (i.e.\ with the same
field points and momenta) in each term of a given Ward identity.

Finally, we comment on the limit of a large Higgs-boson mass in the SM
which is most conveniently discussed in the framework of the non-linear
realization. 
Since in this formulation the scalar self interactions, which become
strong for a heavy Higgs boson, are independent of the would-be 
Goldstone-boson 
fields, 
the SM formally reduces to the so-called {\it gauged non-linear
$\si$-model\/} (GNLSM)%
\footnote{The difference between the GNLSM and the SM for $\MH\to\infty$
is of $\O(\MH^{-2})$ at tree level. However, already at one-loop order
differences of $\O(\log\MH)$ and $\O(1)$ exist, which can be quantified
by an effective Lagrangian \cite{sdcgk,hemo}.}%
\cite{gnlsm}.
The Lagrangian of the (non-renormalizable) GNLSM follows from the SM
Lagrangian with non-linear scalar realization
simply by disregarding the physical Higgs field $H$. Thus,
omitting all terms (and remarks) involving $H$ in this
section, the results for the SM with non-linearly realized scalar sector
transfer to the GNLSM. In particular, the basic Ward identities 
\refeq{eq:WIET} and the construction of physical gauge-boson fields
remain valid. 
However, since the GNLSM does not respect unitarity, the terms
originating from $v_a^\mu$ in \refeq{eq:ET0} do in general not vanish at
high energies.
The range of validity of the ET \refeq{eq:ET} is modified.
This range can, for instance, be determined by applying power counting to
the single Feynman graphs
as proposed in \citere{Gr95b} both for the
GNLSM as well as for the SM with non-linear scalar realization and
arbitrary $\MH$.

\section{SUMMARY AND CONCLUSIONS}
\label{se:sum}

For the conventional approach for quantizing gauge fields,
it is a well-known fact that Dyson summation in general spoils the 
underlying gauge symmetry in finite orders of perturbation theory, 
i.e.\ at the level of Green functions or 
$S$-matrix elements the Ward identities are violated.
Consequently, gauge cancelations for high-energetic longitudinal gauge
bosons, and in particular the validity of the 
Goldstone-boson equivalence theorem (ET),
in general are disturbed when Dyson resummation is
applied.
We have explicitly derived and analyzed the Ward identities for connected 
Green functions within the background-field method (BFM) and found that 
the above-mentioned problems do not occur in this approach. It turns out that
{\em BFM Ward identities for connected Green functions are exactly valid 
loop order by loop order in perturbation theory even after Dyson resummation}
if the inverse propagators, \ie the two-point vertex functions,
are evaluated in the same 
loop order as all other vertex functions. 
In the same way {\em the ET within the BFM is valid after Dyson summation.}

As frequently discussed in the literature, within the conventional
formalism the formulation of the ET in higher orders is non-trivial and 
requires the inclusion of correction factors 
which depend on the renormalization scheme of the unphysical sector.
In view of this, we have analyzed the ET within the BFM in
detail. Starting from the gauge invariance of the background-field
effective action, we 
have derived Ward identities for on-shell amputated Green functions 
with arbitrary insertions of gauge-fixing terms. Using these Ward identities,
which are formally equivalent to those in the conventional formalism, and a 
careful amputation procedure, we have derived the ET. 
The BFM simplifies, in particular, the amputation procedure and thus the
precise formulation of the ET in higher orders.
The correction factors that modify the na\"\i ve form of the ET are given by 
the residues of the gauge-boson propagators which are needed for the 
calculation of the $S$-matrix elements anyhow.

We have argued that our formulation of the ET is independent of the 
parameterization of the Higgs sector. 
This has been explicitly confirmed for 
a non-linear realization of the Higgs sector.
In this context, 
we have also discussed the validity of the ET in the heavy Higgs-boson limit
of the SM and the gauged non-linear 
$\si$-model.
The power-counting methods needed in these cases to assess the range of 
validity of the ET can be directly transferred from the conventional 
formalism to the BFM.

As in previous applications, the BFM 
turns out to be superior to the conventional formalism.
The above-mentioned advantages can be traced back to the gauge
invariance of the background-field
effective action and the associated Ward identities.

\section*{ACKNOWLEDGMENT}

We thank H.~Spiesberger and G.~Weiglein for a careful reading of 
the manuscript.
We are indebted to H.-J.~He for useful comments on the
determination of the correction factors of the equivalence 
theorem in the conventional formalism \cite{He92}
and for drawing our attention to \citere{He95}.

\appendix
\section*{APPENDIX}

\section{MOMENTUM-SPACE AND FIELD CONVENTIONS}
\label{app:momconv}

In order to avoid confusion, we summarize our conventions for the
Fourier transformation needed for the transition from coordinate to
momentum space. In this section all fields are generically denoted by
$\phi$. We start by defining
the Fourier transform of a general vertex function:
\beq
\Ga^{\phi_1\cdots\phi_n}(x_1,\dots,x_n) = 
\int\frac{\d^4k_1}{(2\pi)^4}\cdots\int\frac{\d^4k_n}{(2\pi)^4}\,
\exp\biggl\{\ri \sum_{l=1}^n k_lx_l\biggr\}\, 
\tilde\Ga^{\phi_1\cdots\phi_n}(k_1,\dots,k_n).
\eeq
In $\tilde\Ga^{\phi_1\cdots\phi_n}(k_1,\dots,k_n)$ all momenta are
incoming, and usually a $\de$-function for total momentum conservation
is split off,
\beq
\tilde\Ga^{\phi_1\cdots\phi_n}(k_1,\dots,k_n) =
(2\pi)^4\de^{(4)}\biggl(\sum_l k_l\biggr)\,
\Ga^{\phi_1\cdots\phi_n}(k_1,\dots,k_n).
\eeq
Transforming the fields via
\beq
\phi_l(x) = \int\frac{\d^4k}{(2\pi)^4}\,
\exp\left\{\ri k_lx_l\right\}\,\phi_l(k),
\eeq
the generating functional for the vertex functions, the effective
action, possesses the two representations
\beqar
\Ga[\phi] &=& \sum_n\,\frac{1}{n!}\,\sum_{\phi_1,\dots,\phi_n}\,
   \int\d^4x_1\,\phi_1(x_1)\cdots\int\d^4x_n\,\phi_n(x_n)\,
   \Ga^{\phi_1\cdots\phi_n}(x_1,\dots,x_n), \nn\\
&=& \sum_n\,\frac{1}{n!}\,\sum_{\phi_1,\dots,\phi_n}\,
   \int\frac{\d^4k_1}{(2\pi)^4}\,\phi_1(-k_1)\cdots   
   \int\frac{\d^4k_n}{(2\pi)^4}\,\phi_n(-k_n)\,
   \tilde\Ga^{\phi_1\cdots\phi_n}(k_1,\dots,k_n).
\hspace{2em}
\eeqar
Because of the different integration measures $\int\d^4x$ and
$\int\d^4k/(2\pi)^4$ it is natural to define the functional derivatives
to extract the vertex functions in coordinate and momentum space as
\beq
\frac{\de}{\de f(x)}f(y) = \de^{(4)}(x-y), \qquad
\frac{\de}{\de f(p)}f(k) = (2\pi)^4\de^{(4)}(p-k).
\eeq

The transition to connected Green functions is performed via the
Legendre transformation
\beq
\ri J_{\phi_l^\dagger}(x) = -\frac{\de\Ga}{\de\phi_l(x)}, \quad
\phi_l(x) = \frac{\de\Zc}{\ri\de J_{\phi_l^\dagger}(x)},  \quad
\Zc[J_\phi] = \Ga[\phi] + \ri\sum_l\int\d^4x\,J_{\phi^\dagger_l}(x)\phi_l(x).
\eeq
Transforming the sources via
\beq
J_{\phi_l}(x) = \int\frac{\d^4k}{(2\pi)^4}\,
\exp\left\{\ri k_lx_l\right\}\,J_{\phi_l}(k),
\eeq
the Legendre transformation takes the following form in momentum space:
\beq
\ri J_{\phi^\dagger_l}(-k) = -\frac{\de\Ga}{\de\phi_l(k)}, \quad
\phi_l(k) = \dzd{J_{\phi^\dagger_l}(-k)},  \quad
\Zc[J_\phi] = \Ga[\phi] + \ri\sum_l\int\frac{\d^4k}{(2\pi)^4}\,
   J_{\phi^\dagger_l}(-k)\phi_l(k).
\eeq
This leads to the two representations for the generating functional $\Zc$
\beqar
\Zc[\phi] &=& \sum_n\,\frac{\ri^n}{n!}\,\sum_{\phi_1,\dots,\phi_n}\,
    \int\d^4x_1\,J_{\phi_1}(x_1)\cdots\int\d^4x_n\,J_{\phi_n}(x_n)\,
    G^{\phi_1\cdots\phi_n}(x_1,\dots,x_n), \nn\\
&=& \sum_n\,\frac{\ri^n}{n!}\,\sum_{\phi_1,\dots,\phi_n}\, 
    \int\frac{\d^4k_1}{(2\pi)^4}\,J_{\phi_1}(-k_1)\cdots
    \int\frac{\d^4k_n}{(2\pi)^4}\,J_{\phi_n}(-k_n)\,
    \tilde G^{\phi_1\cdots\phi_2}(k_1,\dots,k_n).
\hspace{2em}
\eeqar
Consequently, the {\em connected} Green functions
$G^{\phi_1\cdots\phi_n}(x_1,\dots,x_n)$ transform via
\beq
G^{\phi_1\cdots\phi_n}(x_1,\dots,x_n) = 
\int\frac{\d^4k_1}{(2\pi)^4}\cdots\int\frac{\d^4k_n}{(2\pi)^4}\,
\exp\biggl\{\ri \sum_{l=1}^n k_lx_l\biggr\}\, 
\tilde G^{\phi_1\cdots\phi_n}(k_1,\dots,k_n),
\eeq
where the momenta $k_l$ are incoming. Again usually the $\de$-function
for total momentum conservation is split off,
\beq
\tilde G^{\phi_1\cdots\phi_n}(k_1,\dots,k_n) =
(2\pi)^4\de^{(4)}\biggl(\sum_l k_l\biggr)\,
G^{\phi_1\cdots\phi_n}(k_1,\dots,k_n).
\eeq

In \refeq{eq:WIZ} both explicit sources $J_{\phi_l}(x)$ and the
corresponding derivatives $\de/\de J_{\phi_l}(x)$ occur. 
The transition to momentum space is performed by applying
$\int\d^4x\,\exp\left\{-\ri kx\right\}\cdots$ to \refeq{eq:WIZ}, 
where $k$ is the momentum flowing into the considered field point $x$. 
This transforms the different terms as follows
\beqar
\int\d^4x\,\exp\left\{-\ri kx\right\}\,\dzd{J_\phi(x)} &=& 
   \dzd{J_\phi(-k)}, \\
\int\d^4x\,\exp\left\{-\ri kx\right\}\,J_{\phi_1}(x)\dzd{J_{\phi_2}(x)} &=&
   \int \frac{\d^4q}{(2\pi)^4}\,J_{\phi_1}(q)\dzd{J_{\phi_2}(q-k)},
   \label{eq:JdZdJ} \\
\int\d^4x\,\exp\left\{-\ri kx\right\}\,
   \dzd{J_{\phi_1}(x)}\dzd{J_{\phi_2}(x)} &=&
   \int \frac{\d^4q}{(2\pi)^4}\,\dzd{J_{\phi_1}(q-k)}\dzd{J_{\phi_2}(-q)}.
   \label{eq:dZdJ2}
\eeqar

The fields that label the vertex and Green functions are defined to be
incoming. For clarity, we list
the relations between two-point functions and propagators:
\beqar
\Ga^{\phi\phi^\dagger}(x,y) = \frac{\de^2\Ga}{\de\phi(x)\de\phi^\dagger(y)}, \nn\\
G^{\phi\phi^\dagger}(x,y) = 
\frac{\de^2\Zc}{\ri\de J_\phi(x)\ri\de J_{\phi^\dagger}(y)} &=&
\langle 0|{\cal T}\phi^\dagger(x)\phi(y)|0\rangle_\conn, \nn\\
\int \d^4 y \, \Ga^{\phi\phi^\dagger}(x,y) \, G^{\phi\phi^\dagger}(y,z) &=&
-\de(x-z), \nn\\
\Ga^{\phi\phi^\dagger}(k,-k) \, G^{\phi\phi^\dagger}(k,-k) &=& -1,
\eeqar
where the field $\phi$ is assumed to mix with no other fields.

\section{PROOF OF THE WARD IDENTITIES FOR ON-SHELL AMPUTATED GREEN
FUNCTIONS}
\label{app:WIproof}

\subsection{Preliminaries}

In order to proof the Ward identities for on-shell amputated Green
functions,
it is useful to rewrite the identities \refeq{eq:WIZ}.
Introducing the operators
\beqar\label{eq:operators}
O_A(x) &=& \ri\partial_x^\mu\dd{\SA^\mu(x)}, \qquad
O_Z(x) = \ri\partial_x^\mu\dd{\SZ^\mu(x)}-\ri\xiZhat\MZ\dd{\Schi(x)}, \nn\\
O_{\pm}(x) &=& \ri\partial_x^\mu\dd{\SWpm^\mu(x)}\mp\xiWhat\MW\dd{\Sphipm(x)},
\hspace{1em}
\eeqar
which appear in all terms that result from the 
gauge-fixing term, we can arrange the Ward identities as
\beqar\label{eq:WIZ2}
\lefteqn{ \frac{1}{\xiAhat}\quabla\Op_A\Zc = 
-\ri\partial\SA
+ e\sum_f Q_f\biggl(\dzd{\Sf}\Sf+\Sfbar\dzd{\Sfbar}\biggr) } 
\qquad \nl
&&{} + \sum_\pm \frac{(\mp e)}{\xiWhat}\dzd{\SWmp^\mu}
       \left(\ri\partial^\mu\Op_\pm\Zc-\xiWhat\SWmp^\mu\right)
     + \sum_\pm e\dzd{\Sphimp} \left(\MW\Op_\pm\Zc\pm\Sphimp\right), 
\nn\\[.5em]
\lefteqn{ \frac{1}{\xiZhat}(\quabla+\xiZhat\MZ^2)\Op_Z\Zc = 
-\ri\partial\SZ-\ri\MZ\Schi } 
\qquad \nl
&&{} - e\sum_f \biggl(\dzd{\Sf}(v_f+a_f\ga_5)\Sf+ 
		      \Sfbar(v_f-a_f\ga_5)\dzd{\Sfbar} \biggr)
     + \frac{\ri e}{2\cw\sw}\dzd{\Schi}{\SH} \nl
&&{} - \frac{e}{2\cw\sw}\dzd{\SH}\left(\MZ\Op_Z\Zc+\ri\Schi\right) 
     - \sum_\pm e\frac{\cw^2-\sw^2}{2\cw\sw}\dzd{\Sphimp} 
       \left(\MW\Op_\pm\Zc\pm\Sphimp\right), \nl
&&{} + \sum_\pm \frac{(\pm e)}{\xiWhat}\frac{\cw}{\sw}\dzd{\SWmp^\mu}
       \left(\ri\partial^\mu\Op_\pm\Zc-\xiWhat\SWmp^\mu\right),
\nn\\[.5em]
\lefteqn{ \frac{1}{\xiWhat}(\quabla+\xiWhat\MW^2)\Op_\pm\Zc = 
-\ri\partial\SWmp\mp\MW\Sphimp } \qquad \nl
&&{} - \frac{e}{\sqrt{2}\sw} \sum_{(f_+,f_-)} 
       \biggl(\dzd{\Sfpm}\frac{1+\ga_5}{2}\Sfmp
              +\Sfbarpm\frac{1-\ga_5}{2}\dzd{\Sfbarmp}\biggr)
     \pm \frac{e}{2\sw}\dzd{\Sphipm}\SH \nl
&&{} + \frac{e}{2\sw}\dzd{\Sphipm}\left(\MZ\Op_Z\Zc+\ri\Schi\right)
     - \frac{e}{2\sw}\biggl(\dzd{\SH}\pm\ri\dzd{\Schi}\biggr) 
       \left(\MW\Op_\pm\Zc\pm\Sphimp\right) \nl
&&{} \mp \frac{e}{\xiAhat}\dzd{\SWpm^\mu}
          \left(\ri\partial^\mu\Op_A\Zc-\xiAhat\SA^\mu\right)
     \pm \frac{e}{\xiZhat}\frac{\cw}{\sw}\dzd{\SWpm^\mu}
         \left(\ri\partial^\mu\Op_Z\Zc-\xiZhat\SZ^\mu\right) \nl
&&{} \pm \frac{e}{\xiWhat}
         \biggl(\dzd{\SA^\mu}-\frac{\cw}{\sw}\dzd{\SZ^\mu}\biggr)
         \left(\ri\partial^\mu\Op_\pm\Zc-\xiWhat\SWmp^\mu\right).
\eeqar
The left-hand sides of \refeq{eq:WIZ2} represent the $\de\Zc/\de J$
terms of \refeq{eq:WIZ}; on the right-hand side the $(\de\Zc/\de J)^2$
terms are combined with the $J_X\de\Zc/\de J$ terms with $X$ denoting a
gauge or a would-be 
Goldstone-boson
field. 

For convenience we write \refeq{eq:WIZ2} generically,
\def\Jterms#1{\langle J_{f,\bar f,\Hhat}\de^{#1}\Zc\rangle}
\def\Zterms#1{\langle \de^{#1}\Zc\rangle}
\def\Opowers#1{\langle \Op^{#1}\rangle}
\beqar\label{eq:WIZ3}
\lefteqn{ \frac{1}{\xihat_a}(\quabla+\xihat_a M_a^2)\Op_a\Zc = 
-\ri\partial J_{\Vhatdag_a}+c_a J_{\phihatdag_a}
+ \Jterms{1} } \nn\\
&&{} + \sum_{b} \left[
\Zterms{1}
	\left(M_{b}^2\Op_{b}\Zc-c_{b}J_{\phihatdag_b}\right) 
+\Zterms{1}
	\left(\ri\partial^\mu\Op_b\Zc-\xihat_{b}J^\mu_{\Vhatdag_b}\right)
\right], \hspace{2em}
\eeqar
with 
$\xihat_\pm=\xiWhat$, $M_\pm = \MW$ and
\beq\label{eq:operator}
O_a(x) = \ri\partial_x^\mu\dd{J_{\Vhat_a}^\mu(x)} 
+ \hat\xi_a c_a \dd{J_{\phihat_a}(x)}.
\eeq
The index 
$a=Z,A,\pm$ refers also to the Ward identity
according to \refeq{eq:WIGammafull}.
Of course, no would-be Goldstone-boson contribution appears for the photon.
The symbol $\Zterms{N}$ represents
terms that do not involve explicit source factors but $N$
derivatives of $\Zc$ with respect to arbitrary sources. 
The symbol $\Jterms{N}$ stands for
terms that involve in addition explicit source factors $J_X$ where 
$X$ is neither a gauge nor a would-be 
Goldstone-boson
field. 
Application of an operator $\Op_{a_l}(x_l)$ to these terms results in
\beq \label{eq:termsprop}
\Op_{a_l}(x_l)\Zterms{N} = \Zterms{N+1}, \qquad
\Op_{a_l}(x_l)\Jterms{N} = \Jterms{N+1},
\eeq
because $\Op_{a_l}(x_l)$ only
contains functional derivatives of gauge and would-be 
Goldstone-boson
fields.

Transforming the operators $O_a(x)$ to momentum space, as specified in
\refapp{app:momconv}, yields
\beq\label{eq:operatormom}
O_a(k) = -k^\mu\dd{J_{\Vhat_a}^\mu(-k)}+\hat\xi_a c_a\dd{J_{\phihat_a}(-k)}.
\eeq

\subsection{Exactly one gauge-fixing term}

We want to show that the Ward identities \refeq{eq:WIZ3} imply
\beq\label{eq:WIZonsgen}
0 = \frac{1}{\xihat_a} (\quabla+\xihat_a M_a^2)O_a(x)\Zc \,+\, \OVT\,,
\eeq
or in momentum space
\beq\label{eq:WIZonsgenmom}
0 = \frac{1}{\xihat_a} (k^2-\xihat_a M_a^2)O_a(k)\Zc \,+\, \OVT\,, 
\eeq
which implies \refeq{eq:WIZonsmom} after
dropping irrelevant factors. Again, 
$\OVT$ (on-shell vanishing terms)
stands for terms that vanish after taking derivatives with
respect to physical fields and subsequent amputation and on-shell
projection.

We consider the Ward identities resulting from \refeq{eq:WIZ3} 
by differentiating with respect to
$n$ physical fields $X_i$, $i=1,\ldots,n\ge2$, with incoming momenta $k_i$
($X_i=\Ahat,\Zhat,\What^\pm,\Hhat,f$).
Their generic structure in momentum space is 
\newcommand{\Diagramleft}[4]{\vcenter{\hbox{%
\begin{picture}(110,60)(-50,-30)
\Line(0,0)(33,18)
\Line(0,0)(33,-18)
\Text(-30,8)[cc]{$\longrightarrow$}
\Line(-45,0)(0,0)           \Text(-30,15)[cb]{#1} \Text(-30,-5)[ct]{#2}
\GCirc(0,0){12}{1}          \Text(40,18)[lc]{#3}  \Text(41,-18)[lc]{#4}
\Vertex(-45,0){2}
\Vertex(33,18){2}
\Vertex(33,-18){2}
\Vertex(25,0){.5}
\Vertex(23,6){.5}
\Vertex(23,-6){.5}
\end{picture}}}
}%
\newcommand{\Diagramleftgauge}[4]{\vcenter{\hbox{%
\begin{picture}(110,60)(-50,-30)
\Line(0,0)(33,18)
\Line(0,0)(33,-18)
\Text(-30,8)[cc]{$\longrightarrow$}
\Photon(-45,0)(0,0){2}{5}   \Text(-30,15)[cb]{#1} \Text(-30,-5)[ct]{#2}
\GCirc(0,0){12}{1}          \Text(40,18)[lc]{#3}  \Text(40,-18)[lc]{#4}
\Vertex(-45,0){2}
\Vertex(33,18){2}
\Vertex(33,-18){2}
\Vertex(25,0){.5}
\Vertex(23,6){.5}
\Vertex(23,-6){.5}
\end{picture}}}
}%
\newcommand{\Diagramleftscalar}[4]{\vcenter{\hbox{%
\begin{picture}(110,60)(-50,-30)
\Line(0,0)(33,18)
\Line(0,0)(33,-18)
\Text(-30,8)[cc]{$\longrightarrow$}
\DashLine(-45,0)(0,0){3}   \Text(-30,15)[cb]{#1} \Text(-30,-5)[ct]{#2}
\GCirc(0,0){12}{1}         \Text(40,18)[lc]{#3}  \Text(40,-18)[lc]{#4}
\Vertex(-45,0){2}
\Vertex(33,18){2}
\Vertex(33,-18){2}
\Vertex(25,0){.5}
\Vertex(23,6){.5}
\Vertex(23,-6){.5}
\end{picture}}}
}%
\newcommand{\Diagramright}[4]{\vcenter{\hbox{%
\begin{picture}(125,60)(-75,-30)
\Line(0,0)(-33,18)
\Line(0,0)(-33,-18)
\Text(30,8)[cc]{$\longleftarrow$}
\Line(45,0)(0,0)            \Text(30,15)[cb]{#1} \Text(30,-5)[ct]{#2}
\GCirc(0,0){12}{1}          \Text(-40,18)[rc]{#3}  \Text(-40,-18)[rc]{#4}
\Vertex(45,0){2}
\Vertex(-33,18){2}
\Vertex(-33,-18){2}
\Vertex(-25,0){.5}
\Vertex(-23,6){.5}
\Vertex(-23,-6){.5}
\end{picture}}}
}%
\begin{eqnarray} \label{fig:WI}
\lefteqn{
  \left[
  k_\mu
  \Diagramleftgauge{$k$}{$\Vhat^\mu_a$}{$X_{1}$}{$X_{n}$}
  -\; \xihat_a c_a
  \Diagramleftscalar{$k$}{$\phihat_a$}{$X_{1}$}{$X_{n}$}
  \right]
  \frac{1}{\xihat_a}(k^2-\xihat_a M_a^2) }\quad   \nl
  &=& 
  \sum_r \sum_{i=1}^n \de_{X_i X_r} d^a_r
  \left.
  \Diagramleft{$k+k_i$}{$X'_r$}{$X_{1}$}{$X_{n}$}
  \right\} \mbox{no } X_i \nl
  &&{}+ \sum_{\mathrm{part.}} \sum_b \sum_{\Fhat} \int\frac{\rd^4q}{(2\pi)^4}
     \, e^a_{b,\hat F}(q)
  \Diagramright{$k-q$}{$\Fhat$}{$X_{{i_{m+1}}}$}{$X_{i_n}$} \nl
  &&\times \underbrace{\left[
  q_\mu
  \Diagramleftgauge{$q$}{$\Vhat^\mu_b$}{$X_{i_1}$}{$X_{i_m}$}
  -\; \xihat_b c_b
  \Diagramleftscalar{$q$}{$\phihat_b$}{$X_{i_1}$}{$X_{i_m}$}
  \right]}_{\mbox{$\longrightarrow$ Ward identity for $\Vhat^\mu_b,\phihat^b$ 
  with $1\le m<n$}} .
  \\[0ex] \nn
\eeqar
The $(n+1)$-point functions in the first line of \refeq{fig:WI} 
correspond to the left-hand side of \refeq{eq:WIZ3}.
The second line of \refeq{fig:WI} results from the terms of the form
$d_r^a J_{X_r}\de\Zc/\de J_{X'_r}$ 
summarized in $\Jterms{1}$ 
in the Ward identities \refeq{eq:WIZ3}, 
where $d_r^a$ denote constant factors.
Because
one of the derivatives 
$\de/\de J_{X_i}(-k_i)$ must act on the explicit
source to produce a non-vanishing term, 
we obtain $n$-point functions
with the incoming field $X_i=X_r$ with momentum $k_i$ converted into the field
$X'_r$ with momentum $k+k_i$.
The $\Zterms{1}O_b\Zc$ terms 
in \refeq{eq:WIZ3} are all of the form 
$\int\d^4q\, e^a_{b,\Fhat}(q)\Bigl(\de\Zc/\de J_{\Fhat}(q-k)\Bigr) 
O_{b}(q)\Zc$ in momentum space,
in particular all of them involve a factor $O_b(q)$.
Differentiating with respect to $n$ external fields 
gives rise to a convolution of Green functions as shown in the
last two lines of \refeq{fig:WI}.
The external fields are distributed in all possible ways over the two
Green functions as indicated by $\sum_{\mathrm{part.}}$. 
The sum over $\Fhat$ runs over all fields that appear
in the combination above and  $e^a_{b,\Fhat}(q)$ is a coefficient that
possibly involves the momentum $q$.
According to \refeq{eq:dZdJ2}, the incoming momentum $k$ 
is distributed to both Green functions.

We now consider the case where
the physical fields $X_i$ are amputated,
put on shell, and multiplied with the corresponding wave functions.
The terms in the second line of \refeq{fig:WI}  
do not possess a pole at $k_i^2=m_{X_i}^2$ and thus vanish after
amputation 
and going on shell with the $X_i$ leg. 
Because all the Green functions in lines 3 and 4 must have at least one 
external 
$X_i$ field (otherwise they vanish owing to $\de\Zc/\de
J_{\Fhat}|_{J_{\Fhat}=0}=0$), 
the number $m$ of $X_i$ legs attached to the Green functions in the
last line must be smaller than the total number $n$ of $X_i$ legs 
but bigger than zero,
\ie $1\le m < n$.
But the last line is just the first line with fewer external legs.
So \refeq{eq:WIZ3} implies that if the first line vanishes for all $m$ with
$1\le m<n$ it vanishes for $n$ as well. 

For $n=2$ the expression in the last line of \refeq{fig:WI}
involves only two-point functions, and the physical fields
$X_i$ are either $\Hhat$, 
$\What^\pm$, $\Ahat$ or $\Zhat$.  
Using \refeq{eq:WIprop},
we get zero for $\Hhat$ and the momentum $k_i^\mu$ for 
a vector field.  After contraction with the corresponding polarization vector
 $\veps_\mu(k_i)$ this vanishes. Thus, the first line of \refeq{fig:WI}
vanishes for $n=2$ and by induction for all $n$.

This proves \refeq{eq:WIZonsgen} 
or equivalently $\langle A|{\cal T}F_a(x)|B\rangle_\conn = 0$.
 
\subsection{General case}

In order to prove 
$\langle A|{\cal T}F_{a_1}(x_1)\cdots F_{a_k}(x_k)|B\rangle_\conn=0$ 
for arbitrarily 
many insertions of gauge-fixing terms, 
or equivalently
\beq\label{eq:multWI}
0 = \left[ \prod_{l=1}^k O_{a_l}(k_l)\right] \Zc \,+\, \OVT \, ,
\eeq
it is necessary to refine our recursive argumentation. For $k>1$ 
the terms originating from $J\de\Zc/\de J$
in general do not vanish separately after amputation and on-shell projection 
but only in combination with $(\de\Zc/\de J)^2$ terms.
This is due to terms that arise when the additional operators
$\Op_{a_l}(k_l)$ act on the explicit source terms in \refeq{eq:WIZ3}.

The Ward identities \refeq{eq:WIZ3} imply
\beqar\label{eq:OaOV}
\lefteqn{ \Op_{a_l}(x_l)\left(M_{a}^2\Op_{a}\Zc-c_{a}J_{\phihatdag_a}\right) =
\Jterms{2}}  \nn\\[.3em]
&\qquad&{} + \sum_{b} \left[
  \left(\Zterms{2} + \Zterms{1} \Op_{a_l}(x_l) \right)
       \left(M_{b}^2\Op_{b}\Zc-c_{b}J_{\phihatdag_b}\right) \right.
\nl &&{}
\qquad + \left. \left(\Zterms{2} + \Zterms{1} \Op_{a_l}(x_l) \right)
       \left(\ri\partial^\mu\Op_{b}\Zc-\xihat_{b}J^\mu_{\Vhatdag_b}\right)
\right], \nn\\[.5em]
\lefteqn{ \Op_{a_l}(x_l)
  \left(\ri\partial^\mu\Op_{a}\Zc-\xihat_{a}J^\mu_{\Vhatdag_a}\right) =
\Jterms{2} } \nn\\[.3em]
&&{} + \sum_{b} \left[
  \left(\Zterms{2} + \Zterms{1} \Op_{a_l}(x_l) \right)
       \left(M_{b}^2\Op_{b}\Zc-c_{b}J_{\phihatdag_b}\right) \right.
\nl &&{}
\qquad + \left. \left(\Zterms{2} + \Zterms{1} \Op_{a_l}(x_l) \right)
       \left(\ri\partial^\mu\Op_{b}\Zc-\xihat_{b}J^\mu_{\Vhatdag_b}\right)
\right],
\eeqar
as can be verified by inserting for $\Op_{a}\Zc$ in the left-hand side
\refeq{eq:WIZ3} and using \refeq{eq:termsprop} and
\beqar
0&=&\Op_{a_l}(x_l)\left(M_{a}^2\xihat_a(\quabla+\xihat_a M_a^2)^{-1}
(-\ri\partial J_{\Vhatdag_a}+c_a J_{\phihatdag_a})-c_{a}J_{\phihatdag_a}\right) ,
\nl
0&=&\Op_{a_l}(x_l)\left(\ri\partial^\mu\xihat_a(\quabla+\xihat_a M_a^2)^{-1}
(-\ri\partial J_{\Vhatdag_a}+c_a J_{\phihatdag_a})-\xihat_{a}J^\mu_{\Vhatdag_a}
\right).
\eeqar
Applying $k-2$ operators $\Op_{a_l}(x_l)$ to \refeq{eq:OaOV} leads to
\beqar\label{eq:WIZ4}
\lefteqn{ \biggl[\prod_{l=2}^k\Op_{a_l}(x_l)\biggr]
        \left(M_{a}^2\Op_{a}\Zc-c_{a}J_{\phihatdag_a}\right) =
\Jterms{k}}  \nn\\[.3em]
&&{} + \sum_{b} \sum_{i=0}^{k-1} \left[
  \Zterms{k-i} \Opowers{i}
       \left(M_{b}^2\Op_{b}\Zc-c_{b}J_{\phihatdag_b}\right)
+ \Zterms{k-i} \Opowers{i}
       \left(\ri\partial^\mu\Op_{b}\Zc-\xihat_{b}J^\mu_{\Vhatdag_b}\right)
\right], \nn\\[.5em]
\lefteqn{ \biggl[\prod_{l=2}^k\Op_{a_l}(x_l)\biggr]
        \left(\ri\partial^\mu\Op_{a}\Zc-\xihat_{a}J^\mu_{\Vhatdag_a}\right) =
\Jterms{k}}  \\[.3em]
&&{} + \sum_{b} \sum_{i=0}^{k-1} \left[
  \Zterms{k-i} \Opowers{i}
       \left(M_{b}^2\Op_{b}\Zc-c_{b}J_{\phihatdag_b}\right)
+ \Zterms{k-i} \Opowers{i}
       \left(\ri\partial^\mu\Op_{b}\Zc-\xihat_{b}J^\mu_{\Vhatdag_b}\right)
\right], \nn
\eeqar
where the symbol $\Opowers{i}$ stands for a product of $i$ operators out of 
$\prod_l\Op_{a_l}(x_l)$.
By taking $n\ge 0$ derivatives $\de/\de J_X$ for arbitrary physical fields $X$ 
and setting all sources to zero one obtains  
Ward identities for Green functions from \refeq{eq:WIZ4}.

In order to prove that the left-hand 
sides of \refeq{eq:WIZ4} are equal to zero
after amputating the physical fields and putting them on shell,
we use induction 
both in the number $n$ of physical fields and in the number
$k$ of operators $\Op_{a_l}$. 
For $k=1$, \ie if the factor $\prod_{l=2}^k \Op_{a_l}(x_l)$ is missing,
\refeq{eq:WIZ4} reduces to \refeq{eq:WIZ3} apart from explicit source
terms $J_{\phihatdag_a},J_{\Vhatdag_a}$, which do not contribute
to on-shell Green functions. Consequently, the left-hand side of
\refeq{eq:WIZ4} vanishes according to \refeq{eq:WIZonsgenmom} for $k=1$.
If we assume that the statement holds for all $k<K$
all terms involving $\Opowers{k}$ with $k<K$ drop out 
in the Ward identity resulting from \refeq{eq:WIZ4} for on-shell
amputated Green functions.
Moreover, the terms $\Jterms{k}$ do not contribute on-shell because they
miss a pole for a physical particle as described in 
the previous section.
Thus, the only terms that can yield non-vanishing contributions obey the 
recursion relations
\beqar\label{eq:WIZ5}
\lefteqn{ \biggl[\prod_{l=2}^K\Op_{a_l}(x_l)\biggr]
        \left(M_{a}^2\Op_{a}\Zc-c_{a}J_{\phihatdag_a}\right) = }
  \nn\\[.3em]
&&{} \sum_{b} \left[
  \Zterms{1} \biggl[\prod_{l=2}^K\Op_{a_l}(x_l)\biggr]
       \left(M_{b}^2\Op_{b}\Zc-c_{b}J_{\phihatdag_b}\right)
+ \Zterms{1} \biggl[\prod_{l=2}^K\Op_{a_l}(x_l)\biggr]
       \left(\ri\partial^\mu\Op_{b}\Zc-\xihat_{b}J^\mu_{\Vhatdag_b}\right)
\right], \nn\\[.5em]
\lefteqn{ \biggl[\prod_{l=2}^K\Op_{a_l}(x_l)\biggr]
        \left(\ri\partial^\mu\Op_{a}\Zc-\xihat_{a}J^\mu_{\Vhatdag_a}\right) = }
  \\[.3em]
&&{} \sum_{b} \left[
  \Zterms{1} \biggl[\prod_{l=2}^K\Op_{a_l}(x_l)\biggr]
       \left(M_{b}^2\Op_{b}\Zc-c_{b}J_{\phihatdag_b}\right)
+ \Zterms{1} \biggl[\prod_{l=2}^K\Op_{a_l}(x_l)\biggr]
       \left(\ri\partial^\mu\Op_{b}\Zc-\xihat_{b}J^\mu_{\Vhatdag_b}\right)
\right]. \nn
\eeqar
In order to show that the left-hand side of \refeq{eq:WIZ5} vanishes
upon taking $n$ derivatives $\de/\de J_X$ with respect to physical
fields $X$ and setting $J=0$ for all fields, we again exploit the recursive 
structure of \refeq{eq:WIZ5} and proceed by induction in $n$. For $n=0$
the statement is trivial owing to
$\Zterms{1}|_{J=0}=0$. Taking $n>0$ derivatives $\de/\de J_X$ from the
right-hand side of \refeq{eq:WIZ5} only those terms can possibly
contribute where at least one derivative is applied to $\Zterms{1}$.
Hence, the statement for $n$ is traced back to the statement for $n-1$,
which completes the induction.
Thus, we have proved
\beqar\label{eq:WIZ6}
 \biggl[\prod_{l=2}^k\Op_{a_l}(x_l)\biggr]
        \left(M_{a}^2\Op_{a}\Zc-c_{a}J_{\phihatdag_a}\right) = \OVT \, , \nl
 \biggl[\prod_{l=2}^k\Op_{a_l}(x_l)\biggr]
        \left(\ri\partial^\mu\Op_{a}\Zc-\xihat_{a}J^\mu_{\Vhatdag_a}\right) =
 \OVT \,.
\eeqar
For $k=2$ and $n=0$ 
we recover the Ward identities for the propagators
\refeq{eq:WIprop}. For $k+n>2$ the explicit source terms drop out and we
obtain \refeq{eq:multWI} from the second equation of \refeq{eq:WIZ6},
since the derivative $\partial^\mu$ translates to a simple momentum factor in
momentum space, which can be dropped.

\section{EXPLICIT ONE-LOOP RESULTS FOR WAVE-FUNCTION NORMALIZATION FACTORS}
\label{app:nomfac}

In \refse{se:physsta} we have described how to calculate $S$-matrix
elements from background-field
Green functions which are renormalized using the
scheme of \citere{bgflong}. For external gauge bosons one needs the
UV-finite 
wave-function renormalization constants $R_{\What}$, $R_{\Ahat\Zhat}$,
and $R_{\Zhat\Zhat}$. These can be determined from the transverse parts
of the  renormalized gauge-boson two-point functions 
(\ses).
In one-loop approximation, the bosonic contributions to the transverse
parts of the BFM self-energies read in the 't~Hooft--Feynman gauge,
i.e.\ for $\xi_Q=1$,
\beqar\label{eq:dsevv}
\hspace{3em} && \hspace{-3.5em}
\left.\Si^{\prime\What^+\What^-}_\rT(\MW^2)\right\vert_{\xi_Q=1}^{\mathrm{bos}}
\; = \; \frac{\alpha}{4\pi\sw^2}\Biggl[ \;
	-\frac{1}{9}
	+\frac{1}{12}\left(\frac{\MH^2}{\MW^2}-1\right)^{\!2}
        B_0(0,\MW,\MH) 
        +\frac{2}{3}\sw^2 B_0(0,0,\MW) 
\nn\\ && {}
        +\frac{\sw^4}{12\cw^4}(1+8\cw^2)B_0(0,\MW,\MZ) 
	-\frac{\MH^2}{12\MW^2}\left(\frac{\MH^2}{\MW^2}-2\right)
	B_0(\MW^2,\MW,\MH) 
\nn\\ && {}
        -8\sw^2B_0(\MW^2,\MW,0) 
        -\frac{1}{12\cw^4}\left(96\cw^6-16\cw^4+6\cw^2+1\right)
        B_0(\MW^2,\MW,\MZ) \nn\\
&&{} -4\sw^2\MW^2B'_0(\MW^2,\MW,\lambda) 
	+\left(\frac{\MH^4}{12\MW^4}-\frac{\MH^2}{3\MW^2}+1\right)
	\MW^2B'_0(\MW^2,\MW,\MH) \nn\\
&&{} -\frac{1}{12\cw^4}(4\cw^2-1)\left(12\cw^4+20\cw^2+1\right)
	\MW^2B'_0(\MW^2,\MW,\MZ) \Biggr] 
	-2\de Z_e-\frac{\cw^2}{\sw^2}\frac{\de\cw^2}{\cw^2}, \nn\\[.5em]
\hspace{3em} && \hspace{-3.5em}
\left.\Si^{\prime\Zhat\Zhat}_\rT(\MZ^2)\right\vert_{\xi_Q=1}^{\mathrm{bos}}
\; = \; \frac{\alpha}{4\pi\cw^2\sw^2}\Biggl[ 
	 \frac{1}{9}(1-2\cw^2)
	+\frac{1}{12}\left(\frac{\MH^2}{\MZ^2}-1\right)^2
	B_0(0,\MZ,\MH) \nn\\
&&{}    -\frac{\MH^2}{12\MZ^2}\left(\frac{\MH^2}{\MZ^2}-2\right)
	B_0(\MZ^2,\MZ,\MH) 
   -\frac{1}{12}\left(84\cw^4+4\cw^2-1\right)B_0(\MZ^2,\MW,\MW)
	\nn\\
&&{} -\frac{1}{12}(4\cw^2-1)\left(12\cw^4+20\cw^2+1\right)
	\MZ^2B'_0(\MZ^2,\MW,\MW) \nn\\
&&{} +\left(\frac{\MH^4}{12\MZ^4}-\frac{\MH^2}{3\MZ^2}+1\right)
	\MZ^2B'_0(\MZ^2,\MZ,\MH) \Biggr] 
	-2\de Z_e-\frac{\cw^2-\sw^2}{\sw^2}\frac{\de\cw^2}{\cw^2}, 
\nn\\[.5em]
\hspace{3em} && \hspace{-3.5em}
\left.\Si^{\Ahat\Zhat}_\rT(\MZ^2)\right\vert_{\xi_Q=1}^{\mathrm{bos}}
\; = \; \frac{\alpha}{4\pi\cw\sw}\MZ^2\Biggl[ 
	\frac{1}{9}
	-\frac{2\cw^2}{3}(6\cw^2-1)B_0(0,\MW,\MW) \nn\\
&&{} +\frac{1}{6}\left(24\cw^4+38\cw^2+1\right)
	B_0(\MZ^2,\MW,\MW) \Biggr] 
	+\frac{\cw}{\sw}\frac{\de\cw^2}{\cw^2}\MZ^2. 
\eeqar
We suppress the 
one-loop fermionic contributions to the 
self-energies since these are identical in the BFM and the conventional
formalism and therefore can be simply inferred from the explicit results 
of \citere{Dehab}. In this reference also the  scalar two-point function
$B_0(p^2,m_0,m_1)$ and its momentum derivative
$B'_0=\partial B_0/\partial p^2$ can be found.
We note that $\Si^{\prime\What^+\What^-}_\rT(\MW^2)$, and thus
$R_{\What}$, gets an IR-divergent contribution contained in
\beq
B'_0(\MW^2,\MW,\lambda) = -\frac{1}{\MW^2}
	\left[1+\log\left(\frac{\lambda}{\MW}\right)\right],
\eeq
where $\lambda$ denotes 
the 
infinitesimal photon mass used as IR regulator.
The counterterms $\de Z_e$ and $\de\cw^2$ read
\beqar
\de Z_e|^{\mathrm{bos}} &=& -\frac{\alpha}{4\pi}\Biggl[
	\frac{7}{2}B_0(0,\MW,\MW)+2\MW^2B'_0(0,\MW,\MW) \Biggr],
\nn\\[.5em]
\de\cw^2|^{\mathrm{bos}} &=& \frac{\alpha}{4\pi}\Biggl[
	-\frac{1}{9}\left(36\cw^4+24\cw^2+1\right)
\nl&&{}
	+\frac{1}{12\sw^2}\left(\frac{\MH^2}{\MZ^2}-1\right)^2B_0(0,\MZ,\MH) 
	-\frac{\cw^2}{12\sw^2}\left(\frac{\MH^2}{\MW^2}-1\right)^2
	B_0(0,\MW,\MH) \nn\\
&& {}+\frac{2\cw^2}{3}\left(6\cw^2+1\right)B_0(0,0,\MW)
     +\frac{1}{12\cw^2}\left(24\cw^4-7\cw^2-1\right)B_0(0,\MZ,\MW) \nl
&& {}-4\cw^2B_0(\MW^2,\MW,0)
	+\frac{\cw^2}{\sw^2}\left(\frac{\MH^4}{12\MW^4}
	-\frac{\MH^2}{3\MW^2}+1\right)B_0(\MW^2,\MW,\MH) \nn\\
&& {}-\frac{1}{\sw^2}\left(\frac{\MH^4}{12\MZ^4}
        -\frac{\MH^2}{3\MZ^2}+1\right)B_0(\MZ^2,\MZ,\MH) \\
&& {}+\frac{4\cw^2-1}{12\sw^2}\left(12\cw^4+20\cw^2+1\right)
	\left(B_0(\MZ^2,\MW,\MW)-\frac{1}{\cw^2}B_0(\MW^2,\MW,\MZ)\right) 
	\Biggr] \nn
\eeqar
in one-loop approximation. While the general background-field gauge-boson
self-energies, and also the derivatives in \refeq{eq:dsevv}, depend on
the quantum
gauge parameter $\xi_Q$, the counterterms $\de Z_e$ and $\de\cw^2$ 
are gauge-parameter-independent \cite{bgflong}. 

Finally, we compare the one-loop expressions for the wave-function
renormalization factors $R$ of the gauge bosons in the linear and
non-linear realization of the Higgs sector. This can easily be done by
inspecting the differences in the Feynman rules. All couplings of exactly 
one would-be Goldstone-boson field to any other fields are identical in both
realizations so that possible differences at one loop could only
originate from quartic couplings between two vector and two scalar
fields. However, also these differences drop out in the factors $R$ so
that the one-loop expressions for the $R$'s coincide in the linear and
non-linear realization. 

We can directly exploit this coincidence
when calculating the leading contributions to the factors $R$ in the
limit of a large Higgs-boson mass. To this end we use the one-loop
effective Lagrangian of \citere{sdcgk} which quantifies the difference 
between the non-linearly realized SM with a heavy Higgs boson and the GNLSM.
At one loop the differences between SM and GNLSM 
in the above-mentioned self-energies read
\beqar\label{MHinfty}
\left.\Si^{\prime\What^+\What^-}_\rT(\MW^2)
\right\vert^{\mathrm{SM-GNLSM}}
&=& \frac{\alpha}{48\pi\sw^2}
11
\left[\De_{\MH}+\frac{5}{6}\right]
\;+\; {\cal O}(\MW^2/\MH^2), \nn\\
\left.\Si^{\prime\Zhat\Zhat}_\rT(\MZ^2)
\right\vert^{\mathrm{SM-GNLSM}}
&=& \frac{\alpha}{48\pi\sw^2}
\left(11-9\frac{\sw^2}{\cw^2}\right)
\left[\De_{\MH}+\frac{5}{6}\right]
\;+\; {\cal O}(\MW^2/\MH^2), \nl
\left.\Si^{\Ahat\Zhat}_\rT(\MZ^2)
\right\vert^{\mathrm{SM-GNLSM}}
&=& -\MZ^2\frac{5\alpha}{24\pi\cw\sw}\left[\De_{\MH}+\frac{5}{6}\right]
\;+\; {\cal O}(\MW^2/\MH^2),
\eeqar
where the $\log\MH$ contributions are contained in the UV-divergent term
$\De_{\MH}$ 
defined by
\beq
\De_{\MH} = \frac{2}{4-D}-\log\left(\frac{\MH^2}{\mu^2}\right)
-\ga_{\mathrm{E}}+\log(4\pi).
\eeq
We have explicitly checked that the $\log\MH$ terms of \refeq{MHinfty} are in
agreement with those obtained by a large-$\MH$ expansion of
\refeq{eq:dsevv}.

\def\refname{REFERENCES}

\end{document}